\documentclass[12pt,preprint]{aastex}
\usepackage{graphics,graphicx,epsfig,lscape}
\usepackage{amssymb}
\newcommand{\lsun}{L$_{\odot}$ }

\newcommand{\eg}{\textit{e.g.}~}

\newcommand{\GALEX}{\textit{GALEX}}
\newcommand{\Spitzer}{\textit{Spitzer}}

\begin{document}

\title{The Young and the Dustless: Interpreting Radio Observations of UltraViolet Luminous Galaxies}

\author{
Antara R. Basu-Zych\altaffilmark{1},
David Schiminovich\altaffilmark{1},
Benjamin D. Johnson\altaffilmark{1},
Charles Hoopes\altaffilmark{6},
Roderik Overzier\altaffilmark{3},
Marie A. Treyer\altaffilmark{2,5},
Timothy M. Heckman\altaffilmark{3},
Tom A. Barlow\altaffilmark{2},
Luciana Bianchi\altaffilmark{7},
Tim Conrow\altaffilmark{2},
Jose Donas\altaffilmark{5},
Karl G. Forster\altaffilmark{2},
Peter G. Friedman\altaffilmark{2},
Young-Wook Lee\altaffilmark{8},
Barry F. Madore\altaffilmark{4},
D. Christopher Martin\altaffilmark{2},
Bruno Milliard\altaffilmark{5},
Patrick Morrissey\altaffilmark{2},
Susan G. Neff\altaffilmark{9},
R. Michael Rich\altaffilmark{10},
Samir Salim\altaffilmark{3},
Mark Seibert\altaffilmark{4},
Todd A. Small\altaffilmark{2},
Alex S. Szalay\altaffilmark{6},
Ted K. Wyder\altaffilmark{2},
Suk Young Yi\altaffilmark{7}}

\altaffiltext{1}{Department of Astronomy, Columbia University, 550 West 120th Street, New York, NY 10027; antara@astro.columbia.edu}
\altaffiltext{2}{California Institute of Technology, MC 405-47, 1200 East California Boulevard, Pasadena, CA 91125}
\altaffiltext{3}{NOAO, Tuscon, Arizona}
\altaffiltext{4}{Observatories of the Carnegie Institution of Washington, 813 Santa Barbara St., Pasadena, CA 91101}
\altaffiltext{5}{Laboratoire d'Astrophysique de Marseille, BP8, Traverse du Siphon, F-13376 Marseille, France}
\altaffiltext{6}{Department of Physics and Astronomy, The Johns Hopkins University, Homewood Campus, Baltimore, MD 21218}
\altaffiltext{7}{Center for Astrophysical Sciences, The Johns Hopkins` University, 3400 N. Charles St., Baltimore, MD 21218}
\altaffiltext{8}{Center for Space Astrophysics, Yonsei University, Seoul 120-749, Korea}
\altaffiltext{9}{Laboratory for Astronomy and Solar Physics, NASA Goddard Space Flight Center, Greenbelt, MD 20771}
\altaffiltext{10}{Department of Physics and Astronomy, University of California, Los Angeles, CA 90095}

\begin{abstract}
Ultraviolet Luminous Galaxies (UVLGs) have been identified as intensely star-forming, nearby galaxies. A subset of these, the supercompact UVLGs, are believed to be local analogs of high redshift Lyman Break Galaxies. Here we investigate the radio continuum properties of this important population for the first time. We have observed 42 supercompact UVLGs with the VLA, all of which have extensive coverage in the UV/optical by \GALEX~and SDSS. Our analysis includes comparison samples of multiwavelength data from the Spitzer First Look Survey and from the SDSS-Galex matched catalogs. In addition we have Spitzer MIPS data for 24 of our galaxies and find that they fall on the radio-FIR correlation of normal star-forming galaxies. We find that our galaxies have lower radio-to-UV ratios and lower Balmer decrements than other local galaxies with similar (high) star formation rates. Optical spectra show they have lower D$_n$(4000) and H$\delta_A$ indices, higher H$\beta$ emission-line equivalents widths, and higher [OIII]5007/H$\beta$ emission-line ratios than normal star forming galaxies. Comparing these results to galaxy spectral evolution models we conclude that supercompact UVLGs are distinguished from normal star forming galaxies firstly by their high specific star formation rates. Moreover, compared to other types of galaxies with similar star formation rates, they have significantly less dust attenuation. In both regards they are similar to Lyman Break Galaxies. This suggests that the process that causes star formation in the supercompact UVLGs differs from other local star forming galaxies, but may be similar to Lyman Break Galaxies.

\end{abstract}


\section{Introduction}

\GALEX~has uncovered a local sample of intensely star-forming galaxies. These galaxies, referred to by \citet{heckman05} as Ultraviolet-Luminous Galaxies (UVLGs), have far-ultraviolet ($\lambda F_{\lambda}$ at 1530$\dot{A}$) luminosities greater than $2 \times 10^{10}$ \lsun. The star formation rates (SFRs) for UVLGs range between 5 to 50 $M_{\odot}$ yr$^{-1}$ \citep{heckman05}, which is $\approx 5-50$ times the SFR for the Milky Way. The most compact, highest surface brightness UVLGs-- the supercompact UVLGS -- have properties similar to those of Lyman Break Galaxies (LBGs) at $z>$3 \citep{heckman05}. \citet{Choopes} has expanded the supercompact UVLG sample to confirm these findings and further determined that supercompact UVLGs also have similar metallicity to LBGs, factors of two or three less than normal galaxies of the same mass. LBGs are named for the technique used to select galaxies at high redshift based on the attenuation of their rest-frame UV continuum shortward of Ly$\alpha$. These were the first high redshift systems discovered by deep optical surveys (\eg, Hubble Deep Field, Hubble Ultra Deep Field), and comprise the most UV luminous galaxies \citep{Steidel95} in the early universe. 

LBGs are very biased at high redshifts \citep{Adelberger05-1}-- suggesting that they may be protogalaxies. Yet, since the exact start of the star formation epoch remains unknown, LBGs might rather describe a later stage in hierarchical structure formation \citep{Steidel99}. Protogalaxies, undergoing a first intense bout of star-formation, ionize surrounding hydrogen gas and emit strongly in Ly$\alpha$ before collapsing farther into fully formed galaxies, one possible explanation for the Lyman Alpha Blobs found near LBGs at $z\sim3.1$ \citep{Steidel00, mark06, Furlanetto}. \cite{Mori} have conducted high resolution hydrodynamic simulations that follow the chemical evolution of primordial galaxies, finding that LBGs resemble infant versions of elliptical and bulge systems in the local universe. At later stages of galaxy evolution, individual protogalaxies are believed to merge and form larger galaxies. Therefore, a better understanding of LBGs is essential for describing galaxy formation and evolution. Since observations at high redshifts (LBGs are found at $z \gtrsim3$) suffer from lower signal-to-noise and surface brightness dimming, the study of local supercompact UVLGs offers an excellent alternative for studying the physics of star formation in these distant protogalaxies.  Lensed LBGs \citep{Smail} can provide complementary approaches to studying the details of LBGs at high redshift.

Separate from the Lyman Break technique, several other techniques have been used to identify intensely star-forming galaxies. As local analogs of LBGs, two classes of galaxies appear to be promising candidates based on their morphologies \citep{Lowenthal05}. Luminous Blue Compact Galaxies, selected by high surface brightness in the I-band \citep{Phillips}, are intermediate-redshift, blue, compact galaxies, with luminosities $\sim$L$^*$ \citep{Garland04, Garland05, Guzman03}; HII galaxies (or Blue Compact Dwarfs), defined by \cite{gildepaz} by high surface brightness, blue colors and low stellar mass, are metal-poor starbursts with an underlying older stellar population. Significant multi-wavelength work has been done to study these local, star-forming galaxies in more detail \citep{guzman05,pg,Rosenberg06}. While some of these samples may overlap with the UVLG sample, the supercompact UVLGs serve uniquely to compare with LBGs because of similar selection methods-- both supercompact UVLGs and LBGs were selected based on UV star formation properties.  Although the UV luminosities have not been studied for Luminous Blue Compact Galaxies yet, HII galaxies are 2 orders of magnitude less luminous in the UV \citep{Choopes}, and therefore have much lower star formation rates.  

Analyzing star formation in galaxies provides essential clues for understanding their structure and evolution. While many star formation indicators exist, each is sensitive to a different subset of the star forming population; the presence of dust, the metallicity, the age or the shape of the star formation history might lead to variations in these separate diagnostics. Short-lived and massive O and B stars predominantly produce ultraviolet radiation. This radiation may be reprocessed by dust surrounding young O and B stars and emitted as infrared radiation. Both ultraviolet and infrared radiation (IR) trace stars formed within 10$-$100 million years. Another excellent tracer of star formation (SF) is the H$\alpha$ produced in the HII regions around young O stars. Sensitive to only the most massive type of stars, the H$\alpha$ lasts for only $\sim$5 million years. As these massive stars die in supernovae, they produce high-energy cosmic rays. The 1.4 GHz continuum radiation is dominated by synchrotron radiation produced by these charged particles accelerated in galactic magnetic fields. Studies by \citet{Kennicutt} and \citet{Schmitt} have compared and correlated these various wavelength regimes, while \citet{Bell03}, \citet{Hopkins} and \citet{Sullivan} include the discussion of timescales and star formation history on SF indicators. 

In this paper, we explore the relationship between the 1.4 GHz radio emission and the far-ultraviolet (FUV) emission for supercompact UVLGs compared to typical galaxies, and incorporate optical spectral parameters and stellar population synthesis models to interpret our observations. Where available, our analysis is supplemented by 70 $\mu$m data. Particularly, we focus on disentangling two effects using our multiwavelength observations: dust attenuation and star formation history. The radio correlates with the total star formation rate, while the ultraviolet measures the attenuated star formation rate \citep{Bell03} -- therefore, the ratio of 1.4GHz SFR to FUV SFR is sensitive to the amount of attenuation. Additionally, the radio signature of star formation appears on a delayed timescale ($\sim 3\times10^{6}- 3\times10^{7}$ years after a single, instantaneous burst) compared to the FUV (within $3\times10^{6}$ years)-- therefore, the offset between UV SFR and radio SFR is also affected by recent star formation history. We use various spectral measures to help separate the effect of dust attenuation from those of star formation history on the radio SFR to UV SFR ratio. We compare the radio luminosities and other properties of supercompact UVLGs with those of LBGs at higher redshifts \citep{Reddy}. In Section 2 we introduce our various samples and describe our data analysis. We discuss our interpretation and include models in Section 3. Finally, we state our conclusions in Section 4. We adopt the following cosmology for all of our calculations: $H_0=70$ km s$^{-1}$ Mpc$^{-1}$,  $\Omega_m= 0.30$ and $\Lambda_0=0.70$. 
 
\section{Data and Analysis}

In this paper, we analyze how our sample of 42 supercompact UVLGs at $0.1<z<0.3$ compares with other star-forming galaxies. We employ an additional three samples that comprise local, ordinary galaxies for the same redshift range.  In creating our samples of comparison galaxies, we draw from the same family of data as the supercompact UVLGs (i.e., UV-\GALEX, Optical-SDSS) to offer the most complementary comparisons with the least introduction of systematic errors that might arise from combining disparate data sets. The comparison samples are believed to include a heterogenous mix of local, star-forming galaxies. In the following sections, we describe each data set separately in detail, and then compare the properties of the entire ensemble of galaxies.

\subsection{Supercompact UVLGs}
\begin{figure}[tbp]
   \vspace{0.0in}
   \centering
   \includegraphics[height=4.5in]{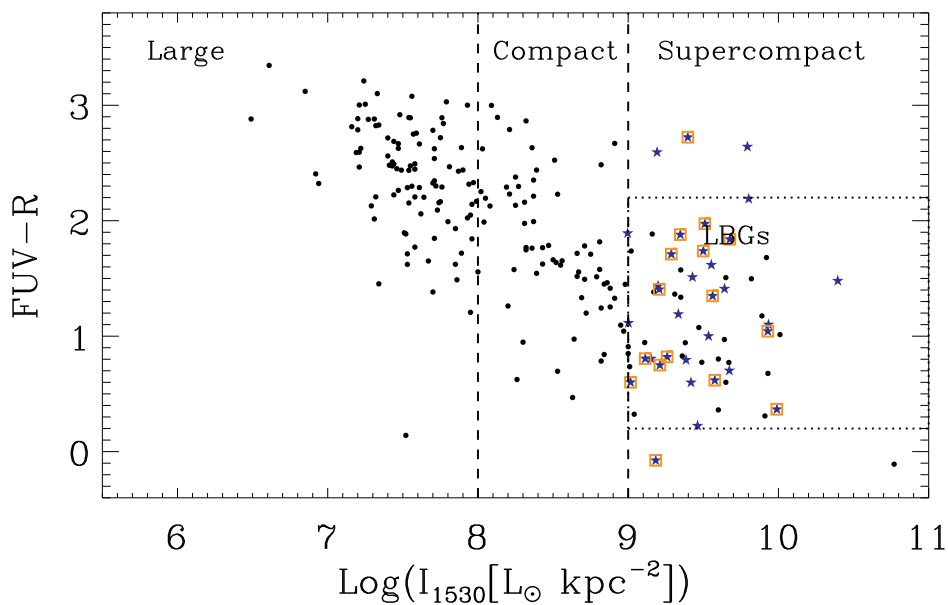} 
   \caption{The blue stars designate the supercompact UVLGs studied in this paper. The black points come from \cite{Choopes} and show how these supercompact UVLGs compare with other UVLGs in FUV surface brightness and  color. We mark the VLA non-detected supercompact UVLGs with orange squares. The dashed vertical lines separate the large from the compact and supercompact UVLGs; the dotted box bounds the color and surface brightness values that describe LBGs. \label{charles}}
\end{figure}

UVLGs (galaxies with $L_{FUV} > 10^{10.3} L_{\odot}$) were selected from the \GALEX~(either MIS-- Medium Imaging Survey or AIS-- All Sky Survey \citep{Patrick}) and Sloan Digital Sky Survey (SDSS) DR2 cross-matched data. \citet{heckman05} and \citet{Choopes} separate UVLGs into three categories: supercompact (I$_{FUV} \geq 10^{9}$ L$_{\odot}$ kpc$^{-2}$), compact ($10^8$ L$_{\odot}$ kpc$^{-2} \leq $I$_{FUV} < 10^9$ L$_{\odot}$ kpc$^{-2}$) and large (I$_{FUV} <10^{8}$ L$_{\odot}$ kpc$^{-2}$), based on their effective radii in SDSS. Supercompact UVLGs show the most similarity with LBGs and are most intensely star forming. Here we discuss 42 supercompact UVLGs observed with the Very Large Array (VLA\footnote[1]{The National Radio Astronomy Observatory is a facility of the National Science Foundation operated under cooperative agreement by Associated Universities, Inc.} ). Fig. \ref{charles} shows our sample of supercompact UVLGs in comparison with other UVLGs from \citet{Choopes}.

Our 1.4GHz continuum observations at the VLA were divided into two programs-- one, in spectral line mode (with 8 channels/IF, each with 3.125 MHz bandwidth), the other in continuum mode. In this second program, we took our data in continuum mode with 50 MHz bandwidth to increase our sensitivity by $\sqrt{2}$. 

For our first program, $\sim 1$ hr B-array observations were taken at 20 cm in spectral-line mode in two IFs, centered at 1.465 GHz and 1.385 GHz, with 25 MHz bandwidth. We observed one of our flux calibrators (`1331+305' (3C286), `0542+498' (3C147), `0137+331' (3C48)) for 10 minutes at the start and end of each of our three observing runs on April 11, April 13, and April 22, 2005. For each source, we applied bandpass, amplitude and phase calibrations using standard Astronomical Image Processing System (AIPS) tasks. We cleaned a giant field of $4096\times 4096$ pixels (2.3 deg$^{2}$) in 1000 iterations, to find the brightest sources whose sidelobes might have contaminated the primary beam. Then we prepared field boxes from the clean components in the previous step, and 3D-cleaned $\sim$ 0.5 deg$^{2}$ images down to 250 $\mu$Jy (near the theoretical 5$\sigma$). The final images have 48 $\mu$Jy $< RMS <$ 78 $\mu$Jy.

The analysis of the second program data was identical (with the exception of bandpass calibration) to the first run. Although our sensitivity was better in general, some fields suffered greatly from noise -- without bandpass calibration, sufficient cleaning of nearby bright sources posed a challenge. Therefore, our RMS spans a large range: 39 $\mu$Jy$< RMS <$ 150 $\mu$Jy. 

In the first run, 28 supercompact UVLGs were observed and 15 were detected. The second run included 15 supercompact UVLGs (8 detections and 7 non-detections) from the first run, providing a consistency check between the two runs.  This second run added 8 new detections, bringing the final tally to 23 detected supercompact UVLGs, and $5\sigma$ upper limits for the remaining 19. We note that the non-detections (shown with orange boxes) do not occupy any special location in Fig. \ref{charles}.  

\begin{figure}[tbp]
   \vspace{0.0in}
   \centering
   \includegraphics[width=7in]{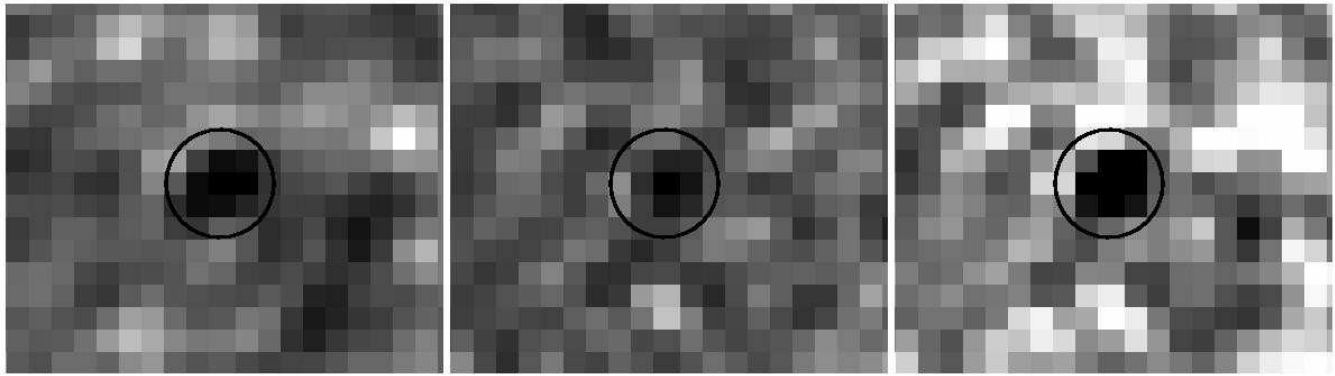}
   \caption{Stack 1, stack 2, and stack\_all (left, center, right panel). The black circle denotes 5$^{\prime\prime}$. These detections measure 141, 157, and 150 $\mu$Jy for stack 1, stack 2 and stack\_all, respectively, implying radio SFRs of $\sim$ 5.8, 4.7, and 5.2  M$_\odot$ yr$^{-1}$.\label{stack}}
\end{figure}

We stacked the non-detections from each run separately (called stack~1, with 12 galaxies and stack~2, with 9 galaxies), as well as combined the non-detections from both of the runs (stack\_all). Some non-detections were omitted in the stacking process, if the field contained nearby bright sources or excessive noise that might have contaminated the stacked image. We obtain detections of 141$\pm$21 $\mu$Jy (stack~1), 157$\pm$18 $\mu$Jy (stack~2) and 149$\pm$15 $\mu$Jy (stack\_all). The stacked images are shown in Fig.  \ref{stack}. Errors were derived from the bootstrapping method. The bootstrapping technique \citep{DasBoot} tests how individual entries affect the stacked average. By generating a random set from the original list, we effectively duplicated some of the values while eliminating others ($\sim 37\%$ of the original entries got replaced by duplicate entries). We calculated a new average. This process was repeated $N Log(N)$ times, where N is the number of samples-- in this case the number of images stacked to produce one stacked image. The standard deviation of these simulated data corresponds to the error in our stacked result. 

The radio emission was k-corrected by $(1+z)^{\alpha}$ with $\alpha= 0.8$, since we assumed a radio spectrum of the form: $S_\nu\propto\nu^{-0.8}$. We consistently use the integrated radio intensities throughout this paper, unless explicitly stated otherwise. We derived the FUV luminosities from magnitudes given in the \GALEX~catalog. For  cases where \GALEX~observed one supercompact UVLG on multiple visits, we computed an exposure-weighted average magnitude. The FUV magnitude was Galactic extinction-corrected by 8.24$\times$E(B$-$V), determined using \cite{wyder}, and k-corrected applying the kcorrect\_4.1.4 IDL routine \citep{kcor} (using the optical u'g'r'i'z' bands from SDSS and FUV, NUV from \GALEX~for spectral fitting). Table 1 summarizes the observations for the supercompact UVLGs. We include 70$\mu$m \Spitzer~data for comparison. We used the 70 $\mu$m mosaic images created by the Spitzer Science Center pipeline and measured the fluxes of point sources at the expected locations of the supercompact UVLGs using 30$^{\prime\prime}$ circular apertures in IRAF. The aperture corrections given on the Spitzer website were applied.

\subsection{Comparison Sample 1: \GALEX~MIS, SDSS DR2 and VLA}
As described above, the VLA fields from our first program were centered on supercompact UVLGs. The spectral line mode in the first program allowed us to obtain 1.4GHz data for serendipitous galaxies in the same fields, since the bandpass calibration made possible the proper cleaning of wider images. After applying the AIPS data reduction tasks, we found a total of 24 other galaxies in these fields that existed in both the SDSS DR2 and \GALEX~GR1 MIS data (separate from the 16 supercompact UVLGs), which we believe to represent a sample of star-forming galaxies. We refer to these galaxies as our `MIS/DR2/VLA' set. The stellar mass was derived for these galaxies according to the prescription in \citet{Brinchmann}. The UV, radio and optical properties were determined in the same way as for the supercompact UVLGs. 

\begin{figure}[tbp]
   \vspace{0.0in}
   \centering
   \includegraphics[height=5in]{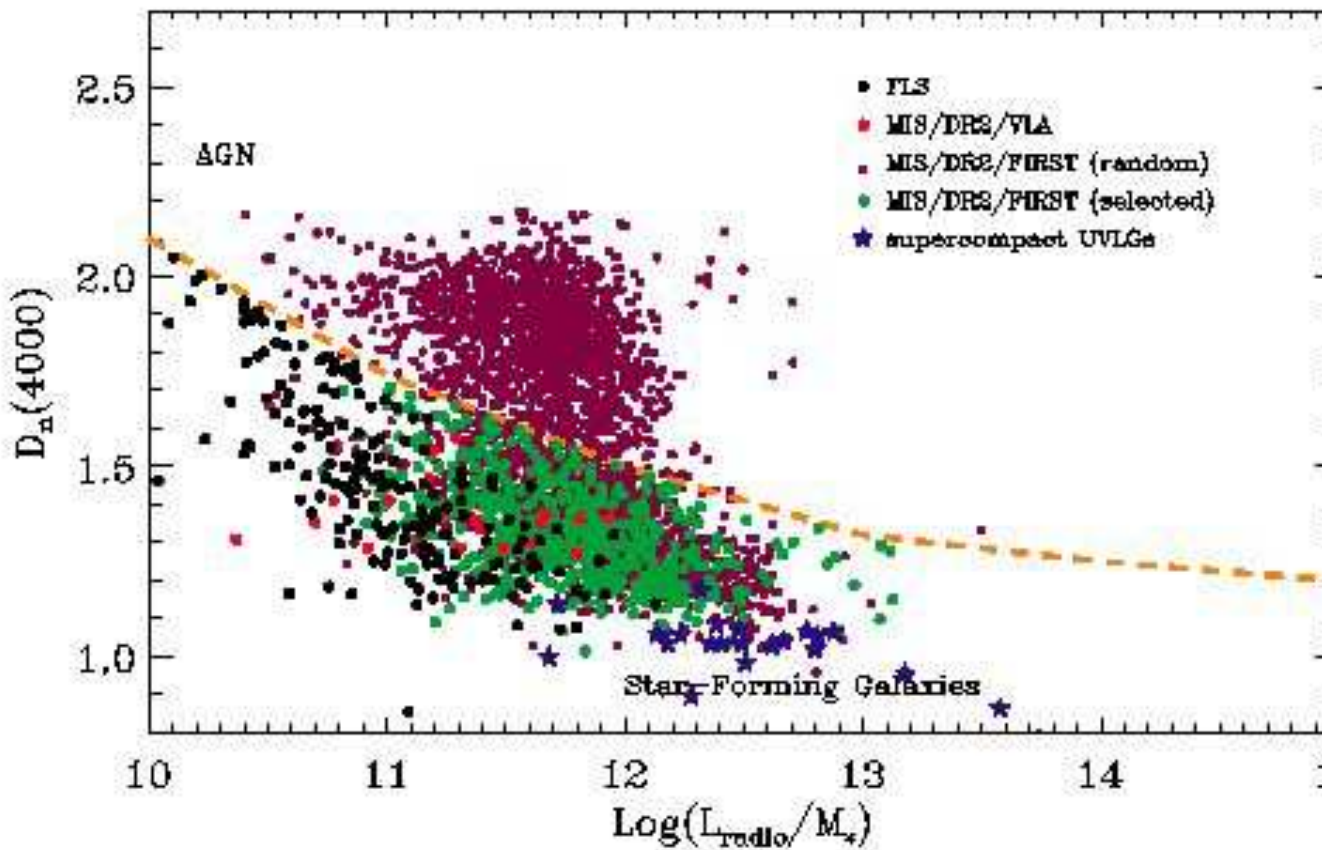} 
   \caption{Using \citet{Best} to apply an AGN cut: the galaxies above the orange line are marked AGN and excluded from our analysis, while the galaxies below the line remain in our samples. We indicate FLS galaxies with black points and the MIS/DR2/VLA sample with red points. The smaller purple points display a random set from the full 60,000 MIS/DR2/FIRST galaxies, whereas the green points show the same data set once AGN have been removed. Supercompact UVLGs appear as blue stars.\label{agncut}}
\end{figure}

Since galaxies containing Active Galactic Nuclei (AGN)-- strongly emitting in UV and radio through processes unrelated to star formation-- could contaminate our assessment of star-forming properties, we eliminated AGN by using the D$_n$(4000) and the radio luminosity per stellar mass as described by \citet{Best} (Fig. \ref{agncut}). As discussed later in this paper, D$_n$(4000) is a common measure of stellar age. Star-forming galaxies with higher D$_n$(4000) contain older stars and lower L$_{radio}/$M$_*$, since L$_{radio}/$M$_*$ gauges the present to past SFR. Therefore radio emission for galaxies above the dotted line in Fig. \ref{agncut}, with high L$_{radio}$/M$_*$, must be dominated by AGN activity rather than star formation. We also note that by this method supercompact UVLGs (marked as blue stars) do not show any evidence of AGN activity, consistent with the selection criterion that \cite{Choopes} employed. 

\subsection{Comparison Sample 2: First Look Survey Galaxies}

The \Spitzer~First Look Survey (FLS) covered 5 square degrees centered at RA= 17h 18m, DEC = +59d 30' (J2000). This region has excellent wavelength coverage-- observations in the infrared by {\it The Spitzer Space Telescope}, radio by the VLA, optical by SDSS, and ultraviolet by  \GALEX. In this study, we restricted our analysis to the SDSS, radio and ultraviolet data. We considered only the sources that have FUV and radio detections. We applied the process described above to remove galaxies dominated by AGN activity, $\sim$30\% of our sample. After these steps, our sample contains FUV and radio luminosities for 189 galaxies. 

The \GALEX~observations were $\sim 20$ks and achieved Far-Ultraviolet (FUV) magnitudes $\leq$ 24.7. VLA observed the FLS galaxies in B array at 20 cm, achieving sensitivities S $> $115 $\mu$ Jy  (see \cite{condon} for full details regarding the VLA observations). 

\subsection{Comparison Sample 3: \GALEX~MIS, SDSS DR2 and FIRST}

The matched catalogs from SDSS and the \GALEX~Internal Release 1.1 MIS comprise data for $\sim$ 45,000 galaxies. As described for the MIS/DR2/VLA galaxies, we eliminated the AGN, and $\sim$ 21,000 galaxies remained in our sample. Faint Images of the Radio Sky at Twenty-cm (FIRST) has observed most of these galaxies in brief 3-minute VLA snapshots in B-array \citep{first}. Very few ($< 3\%$) of the of MIS/DR2 galaxies were detected by FIRST. However, we stacked the FIRST observations to simulate deeper observations of non-radio selected galaxies (like the supercompact UVLGs, and therefore providing an important comparison). 

To perform our stacking, we separated the galaxies into bins by FUV luminosity (from 17.5$<$ log L$_{FUV} <$ 22.5, in increments of 0.1), redshift (from $0.05<z<0.35$ in increments of 0.05), and attenuation ($0.0 < \tau$\footnote[2]{$\tau \equiv \mathrm{ln} [\frac{H \alpha}{H \beta} \frac{1}{2.88}]$ \citep{calzetti94}. Our use of $\tau$ for measuring attenuation is discussed further in the discussion. \label{fn:taufoot}} $<$ 1.0). We calculated a stacked image by taking the mean of the images for each bin, provided that the bin contained at least 30 galaxies. These stacked ``detections'', the henceforward `MIS/DR2/FIRST' galaxies, provide an average measurement of the radio luminosity for typical MIS/DR2 galaxies. In this analysis, the radio intensities are the peak intensities. We assumed that the galaxies are unresolved and the peak intensity could be approximated as the integrated intensity for the stacked galaxies. This might lead to an underestimate in the radio luminosity for the nearby galaxies; FIRST's resolution is 5$^{\prime\prime}$, implying that galaxies larger than 11 kpc at $z=0.1$ would be resolved, while galaxies at $z=0.3$ would remain unresolved at sizes as large as 38 kpc.   

A small number of galaxies ($\sim 2.5\%$) in this `MIS/DR2/FIRST' sample were detected by FIRST. We included these detections in calculating the average radio intensity for the relevant redshift and L$_{FUV}$ bin. However, the detected sources were excluded from the stacking process. We applied the correction, prescribed in \cite{white}, to account for the snapshot bias of unknown origin. This snapshot bias leads to an underestimate of the source flux (even for stacked sources) because the non-linear cleaning process redistributes flux from sources to the background. To estimate errors in our stacked image, we used the bootstrapping method, described earlier. Additionally, we consider the detections as a separate comparison sample, referred to as the `MIS/DR2/FIRST (det)' galaxies. 

\subsection{Computing Radio and UV SFRs}
SFRs were derived from the FUV and radio luminosities using the relations of \citet{Hopkins}. The FUV SFR derivation agrees with that of \cite{Kennicutt} and \cite{Sullivan}. The radio SFR relation matches the derivation given in \cite{Bell03}, which normalizes the radio SFR to match the IR SFR for high luminosity galaxies. We note that this conversion to the radio SFR differs from that in \cite{condon92} and \cite{Sullivan} by a factor of 2 for L$_{1.4GHz} > 6.4\times 10 ^{21}$ W Hz$^{-1}$ (see discussions in \cite{Bell03} and \cite{Hopkins}). 

\citet{Hopkins} used the \cite{Salpeter} Initial Mass Function (IMF) [$\Psi(M) \sim M^{-2.35}$], and a mass range of $0.1 < M_{\odot} < 100$. In order to adopt the \cite{Kroupa} IMF, we divided the SFR conversions given in \citet{Hopkins} by 1.5. The UV and radio SFRs that we used were: 

\begin{equation}
\mathrm{SFR_{FUV}}(\mathrm{M}_{\odot}\mathrm{yr}^{-1})=\frac{\mathrm{L}_{\mathrm{FUV}}}{1.07\times 10^{21}}
\end{equation}

\begin{equation}
\mathrm{SFR_{1.4GHz}}(\mathrm{M_{\odot}~yr^{-1}})= \left\{ \begin{array}{lr}
 \frac{\mathrm{L}_{1.4\mathrm{GHz}}}{2.72 \times 10^{21}} ,& \mbox (\mathrm{L}_{1.4\mathrm{GHz}} > \mathrm{L}_{c})\\
\frac{\mathrm{L}_{1.4\mathrm{GHz}}}{2.72 \times 10^{21}[0.1+0.9 (\mathrm{L}_{1.4 \mathrm{ GHz}}/\mathrm{L}_{c})^{0.3}]}, & \mbox (\mathrm{L}_{1.4GHz} \leq \mathrm{L}_{c}) \end{array}\right. 
\end{equation}

where L$_{c}=6.4\times 10^{21}$ W Hz$^{-1}$, and L$_{\mathrm{FUV}}$ and L$_{\mathrm{1.4GHz}}$ are in W Hz$^{-1}$. We used Eqns 1 and 2 to derive all the UV and radio SFRs throughout this paper. SDSS optical spectral properties have been computed and made available on http://www.mpa-garching.mpg.de/SDSS/. For our analysis in this paper, we used the optical line fluxes (described in \citet{Tremonti}), stellar masses, H$\delta$ absorption line (H$\delta_{A}$) and 4000$\mathrm{\AA}$ break index (D$_n$(4000)) (detailed in \citet{Brinchmann}) from these tables. We note that the UV SFRs and luminosities throughout this paper (for all the samples) are uncorrected for dust attenuation. 

\subsection{The Ensemble}
\begin{figure}[tbp]
   \vspace{0.0in}
   \centering
   \includegraphics[height=6.7in, angle=90]{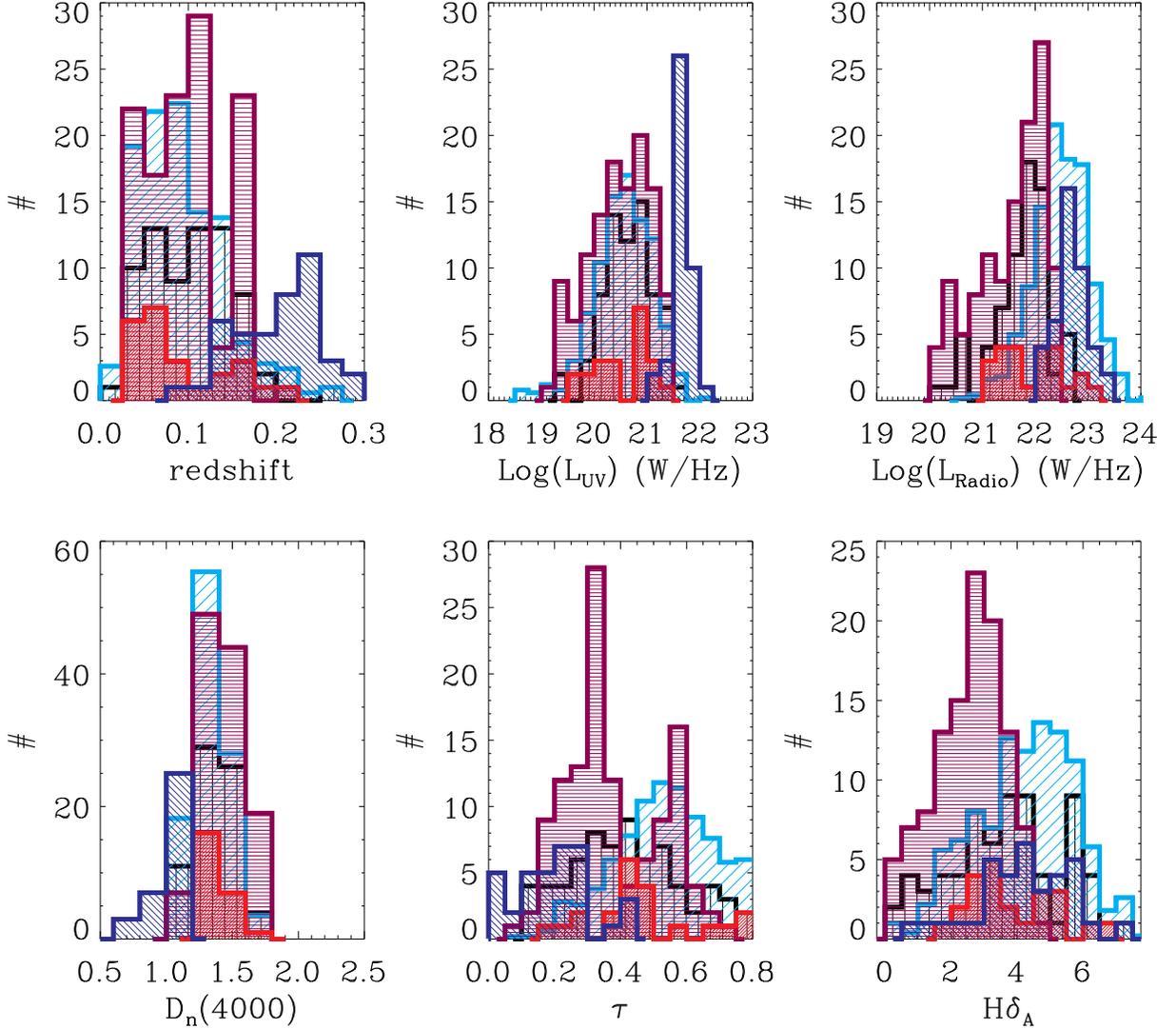} 
   \caption{Comparing redshifts, UV luminosities, radio luminosities, D$_{n}$(4000), $\tau$ (See footnote \ref{fn:taufoot}), and H$\delta_{A}$ for the samples (we show the total number for each sample in parentheses) -- Blue, negatively-sloped diagonal stripes describe supercompact UVLGs (42); violet, horizontal stripes show the {\em stacked} MIS/DR2/FIRST sample (99); cyan, positive-sloped diagonal stripes display MIS/DR2/FIRST detections(526, we scaled this distribution down by a factor of 5 in order to display all the data sets clearly); red, positively-sloped diagonal stripes display MIS/DR2/VLA galaxies (27); black, vertical stripes illustrate FLS galaxies (79).\label{hist}}
   \end{figure}

In this paper, we focus on how supercompact UVLGs compare with other star-forming galaxies. The FLS, MIS/DR2/FIRST, and MIS/DR2/VLA data constitute our comparison samples. In Table 2 we summarize the separate data sets; Fig. \ref{hist} illustrates how the physical quantities (redshift, UV and radio luminosities, D$_n$(4000), $\tau$, and H$\delta_{A}$) in the samples are distributed. We note the following:
	\begin{itemize}
\item All the comparison samples agree in their distributions for redshifts, and UV and radio luminosities. The median redshifts for the MIS/DR2/VLA, MIS/DR2/FIRST detected and stacked points, and FLS are: 0.07, 0.08, 0.08 and 0.1 (the standard deviation for all samples is 0.05). The median UV luminosity (in Log units) for the same aforementioned data sets are: 20.8$\pm$0.5, 20.6$\pm$0.6, 20.5$\pm$0.6 and 20.7$\pm$0.5;  similarly, the median radio luminosity are: 22.1$\pm$0.6, 22.5$\pm$0.5, 21.7$\pm$0.6 and 21.9$\pm$0.6. The stacked MIS/DR2/FIRST galaxies include less luminous radio and UV galaxies, since the longer effective exposure times allow for deeper observations, while MIS/DR2/VLA and MIS/DR2/FIRST (detected) galaxies correspond to more luminous radio sources.
\item Supercompact UVLGs have a redshift distribution peaking at a slightly higher redshift (their median redshift is 0.21$\pm 0.05$). The distributions of UV and radio luminosities are more sharply peaked and at higher values than the other data sets (median values of UV luminosity and radio luminosity are 21.6$\pm$0.2 and 22.7$\pm$0.3 in Log units). This is not surprising since supercompact UVLGs are restricted to those galaxies with UV luminosities greater than $2\times 10^{10} L_{\odot}$. 
\item The comparison samples, MIS/DR2/VLA, MIS/DR2/FIRST (detected) and FLS, were selected to have detections in both the FUV and the radio. Including radio non-detections might have given a closer comparison to the supercompact UVLGs, which were not radio-selected, and would probe the relations at lower radio luminosities. The stacked non-detections of the MIS/DR2/FIRST galaxies serve this purpose. 

\end{itemize}
In summary, the FLS and stacked MIS/DR2/FIRST data provide the deepest observations in both radio and UV. The FLS, MIS/DR2/VLA and MIS/DR2/FIRST (detected) samples include radio and UV selected galaxies, while the supercompact UVLGs are only UV selected.  In fact, the MIS/DR2/FIRST (detected) sample includes the most radio luminous galaxies, a subset of which may be IR luminous galaxies. We include the stacked MIS/DR2/FIRST galaxies for two reasons: to extend our sample to lower radio luminosities, and to compare the supercompact UVLGs with galaxies that were also not radio selected. Where appropriate, we will provide two versions of our plots: one, excluding the stacked MIS/DR2/FIRST data to provide a less cluttered view, and the other with the full set. 

\section{Discussion}

\begin{figure}[tbp]
   \vspace{0.0in}
   \centering
   \includegraphics[height=5in]{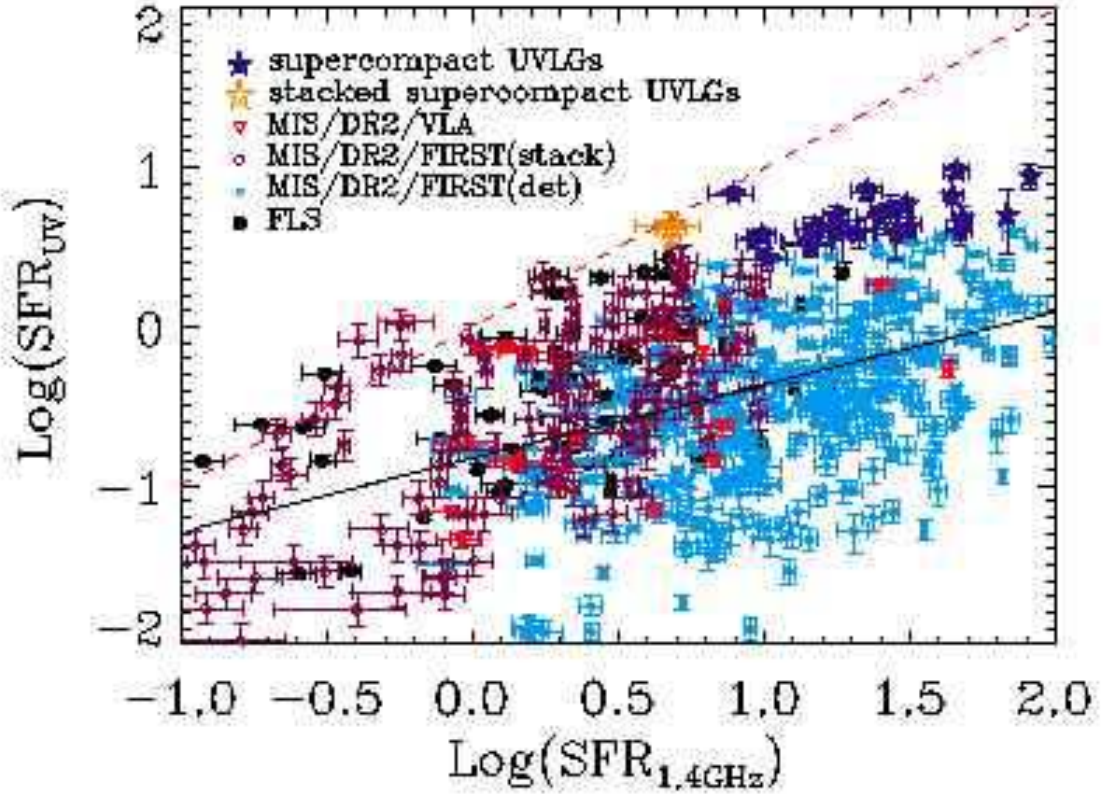} 
   \caption{Comparing the unattenuated SFR ($SFR_{radio}$) with the attenuated SFR ($SFR_{uv}$). Supercompact UVLGs (blue stars) seem to have slightly higher UV emission (or lower radio emission) compared to the other local galaxies (First Look Survey galaxies shown in black dots; MIS/DR2/VLA galaxies are shown as inverted red triangles; stacked MIS/DR2/FIRST galaxies are displayed as violet open circles; detected MIS/DR2/FIRST galaxies are indicated by cyan points.) The solid black line designates the best fit line to all the points (we display this fit in order to guide the eye, without intending to quantify any physical connection between the two axes); the dashed red line marks equivalent radio and UV SFR.\label{raduvsfr}}
\end{figure}

Why are UVLGs so luminous in the ultraviolet? Might they be highly obscured galaxies with high SFRs, that still produce significant UV luminosity? Or do these galaxies have moderate star formation rates, but less attenuation from dust compared to other UV- and optically-selected galaxies? Comparing the radio SFR versus the UV SFR (Fig. \ref{raduvsfr}) shows that supercompact UVLGs have higher star formation in both diagnostics compared to the galaxies in our comparison samples. It should be noted that if we include a sample of IR luminous and ultraluminous galaxies in this figure they would likely fall in the region below the supercompact UVLGs. For all of the galaxies, including the supercompact UVLGs, the UV SFR and radio SFR measures are consistent. The dashed red line shows the line of equivalent star formation rate. We expect that the actual UV SFR underestimates the total star formation rate (here, taken to be the radio SFR) because UV suffers strongly from dust attenuation, if present. The shallower slope of the solid black line (the best fit line to all the data points, including the supercompact UVLGs) compared to the dashed red line suggests that attenuation increases with increasing SFR (similarly, \citet{Wang96} and \citet{Martin05} show that dust extinction increases with both far-infrared(FIR) and FIR$+$UV luminosities for normal late-type galaxies). All the samples of regular galaxies scatter around the fit (black solid line) reasonably well. We provide the solid black line as a visual reference for describing the trend outlined by all the galaxies in our samples but note that the fit is not physically motivated, nor intended to quantitatively relate the radio SFR to the UV SFR. Compared to this trend, the supercompact UVLGs are offset-- these galaxies seem under-luminous in their radio emission compared to the UV. We note, however, that the median radio luminosity for the supercompact UVLGs, L$_{1.4GHz} \sim 5.8 \times 10^{22}$ W Hz$^{-1}$, is consistent with the value determined for LBGs (by stacking LBGs at 1.5$\geq z \geq 3.0$ from the Great Observatories Origins Deep Survey-- North field), L$_{1.4GHz} \sim 5.90$($\pm1.66$) $\times 10^{22}$ W Hz$^{-1}$\citep{Reddy}. 

\begin{figure}[tbp]
   \vspace{0.0in}
   \centering
   \includegraphics[height=5in]{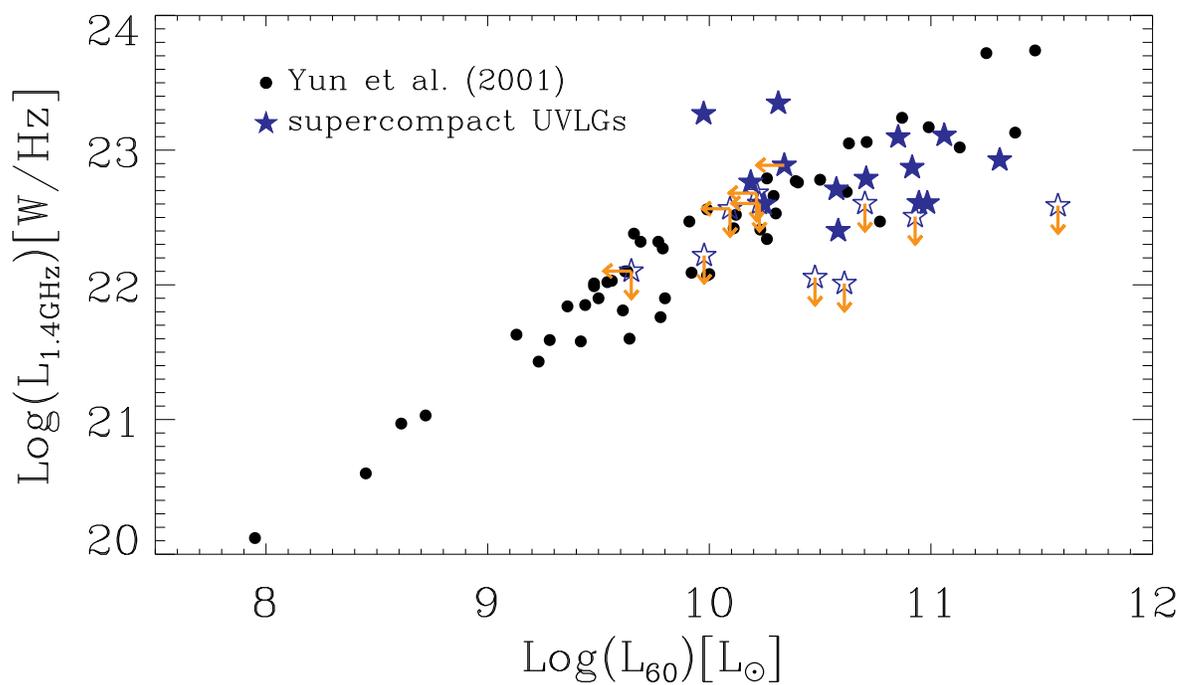} 
   \caption{Radio vs. IR-- comparing the 60 $\mu$m with 1.4 GHz radiation. The supercompact UVLGs follow the points from \cite{Yun}.  Radio or IR non-detections appear as open stars with orange arrows denoting $5\sigma$ upper limits for the 60$\mu m$ (pointing left) and the 1.4GHz (pointing down) data; radio detections are shown by filled stars.\label{radir}} 
\end{figure}

We briefly consider the IR data for supercompact UVLGs -- comparing our L$_{1.4GHz}$ and L$_{60\mu m}$ values with those from \cite{Yun}. We compute the 60$\mu$m rest frame luminosities, L$_{60\mu m} = \nu_{60\mu m}$F$_{60\mu m}$, from the observed 70$\mu$m \Spitzer~data. From Fig. \ref{radir}, we find that the supercompact UVLGs appear to follow the radio-IR correlation. We plan to complete further analysis of the other IR bands to determine the total IR luminosity and IR SFR in a future paper.

Assuming that the radio SFR measures the total SFR we can define L$_{tot, rad}=5.4\times 10^9$ SFR$_{rad}$ \citep{Martin05, Kennicutt}, where L$_{tot,rad}$ is the equivalent bolometric luminosity of young, massive stars as derived from our radio luminosities. Given that the UV SFR underestimates the total star formation proportional to the amount dust attenuation and the SFR derived from the IR makes up this difference, then $SFR_{IR}+SFR_{UV} = SFR_{rad}$. Therefore L$_{IR}+$L$_{UV}=$L$_{tot, rad}$. In the top panel of Fig. \ref{chris}, we plot Log(L$_{tot, rad}$) versus Log($\frac{L_{tot, rad}-L_{UV}}{L_{UV}}$), similar to \citet {Wang96} and \citet{Martin05}. The solid line gives the fit from  \citet{Martin05}, and the dashed line shows the fit from \citet {Wang96}. We see that the samples of normal galaxies have scatter both above and below the fits, but the supercompact UVLGs fall below the relations. In the bottom panel, we plot radio versus UV luminosity, and the lines show the radio luminosity derived from the UV luminosity and the fit from  \citet {Martin05} (solid line) and from \citet {Wang96} (dashed line). Here, we can clearly see that {\em all} the supercompact UVLGs in our sample have radio luminosities lower than expected based on their UV-derived SFR. 

We explore a couple of scenarios which might explain the lower radio to UV emission: first, supercompact UVLGs could be less attenuated; secondly, they could be dominated by younger, less evolved stellar populations compared to other galaxies.

Examining the first option, we determine the dust content of the samples using two independent techniques. The first method uses the ratio of radio to UV radiation from these galaxies. As discussed above, the UV luminosity gives us a measure of the attenuated SFR, while the radio emission can be used to measure the unattenuated SFR. The other technique uses the ratio of line fluxes from H$\alpha$ and H$\beta$: $\tau \equiv$ ln $[\frac{\mathrm{H} \alpha}{\mathrm{H} \beta} \frac{1}{2.88}]$. Both measure the amount of Hydrogen in the system, but H$\beta$ is more sensitive to dust. The natural difference between the line strengths is accounted for by the factor 2.88. Larger values of $\tau$ indicate greater attenuation. Note that the Balmer decrement, or $\tau$, may be a upper limit to the actual amount of attenuation, since emission lines tend to show more attenuation than continuum \citep{Charlot2000}.
 
Fig. \ref{tau} compares the line with the continuum attenuation. To better compare the two attenuation values, we convert both to $\tau_V$ units (the attenuation in the V magnitude), by the following:

\begin{eqnarray}
\frac{F_{H\alpha}}{F_{H\beta}}=\frac{F_{H\alpha, 0}}{F_{H\beta, 0}} e^{-(\tau_{H\alpha}-\tau_{H\beta})}\\
\tau_{H\alpha}=\tau_V \left ( \frac{6563\mathrm{\AA}}{5500\mathrm{\AA}}\right )^{-0.7}\\
\tau_{H\beta}=\tau_V \left (\frac{4861\mathrm{\AA}}{5500\mathrm{\AA}}\right )^{-0.7}
\end{eqnarray}
then, 

\begin{eqnarray}
\nonumber \tau & \equiv  & ln[\frac{F_{H\alpha}}{F_{H\beta}}\frac{1}{2.88}] \\ 
\nonumber &= & ln\left [\frac{F_{H\alpha, 0}}{F_{H\beta, 0}}\frac{1}{2.88}\right ] e^{-(\tau_{H\alpha}-\tau_{H\beta})}\\ 
\nonumber&= & \tau_{H\beta}-\tau_{H\alpha}\\ 
&= & \tau_V/4.84
\end{eqnarray}

and 

\begin{eqnarray}
Ln(SFR_{\mathrm{1.4GHz}}/SFR_{\mathrm{UV}})&=& \tau_V(1500\mathrm{\AA}/5500\mathrm{\AA})^{-0.7} \\
&=&2.48~\tau_V
\end{eqnarray}
Therefore, we define the quantities $\tau_{V, line}$ and $\tau_{V, cont}$ as:
\begin{eqnarray}
\tau_{V, line}& \equiv &4.84\tau\\
\tau_{V, cont}& \equiv &Ln(SFR_{\mathrm{1.4GHz}}/SFR_{\mathrm{UV}})/2.48
\end{eqnarray}

The two diagnostics correlate, although there is significant scatter. The supercompact UVLGs fall farther to the right side of the correlation defined by the other galaxies, towards higher continuum attenuation, and closer to the dashed line (where the line and continuum attenuation are equivalent). Supercompact UVLGs have nearly equal amounts of line and continuum attenuation while other galaxies have higher line attenuation for a given continuum attenuation. 

To compare with higher redshift LBGs: \citet{Reddy} measure continuum attenuation as SFR$_{xray}$/SFR$_{UV}\sim 4.5-5$, which is consistent with our sample's median SFR$_{1.4GHz}$/SFR$_{UV}$ = 4.7.  Additionally, \citet{Erb03} find discrepancy between UV and H$\alpha$ SFR for LBGs at $z>2$, which they attribute to attenuation differences between regions giving rise to continuum versus line emission. 

One explanation for why there is a mismatch between the line and continuum attenuation for the supercompact UVLGs might be the ``birth cloud'' description\footnote[3]{By ``birth cloud'', we refer to the scenario presented by \cite{Charlot2000}. However, we relax their definition of ``birth cloud'' to include stars that may have been formed in earlier episodes, but enclosed within pockets of more recent star formation.}: for young systems, the continuum and line attenuation both arise from the same physical regions: in HII regions surrounding new stars and also the ISM; for older galaxies, the continuum is less attenuated, only by the ISM, while the emission lines are attenuated by both ISM and HII regions. If we assume that most of the stars in supercompact UVLGs are embedded in ``birth cloud'' regions, then we would expect the line and continuum attenuation measures to be similar. Since the continuum attenuation is lower than the line attenuation for other typical galaxies, the supercompact UVLGs would appear to have higher continuum attenuation than the typical galaxy. 

Following this assertion a bit further, we introduce some models. We ran the \citet{BC} stellar population synthesis on three different star formation scenarios (an instantaneous burst, constant 1$M_{\odot}$ yr$^{-1}$, and exponentially declining with $\tau=100$ Myr), for different metallicities (Z=0.008, 0.02, and 0.05 corresponding to Z/Z$_\odot$=0.3,1 and 2.5) and different attenuation amounts, governed by:

\begin{equation}
\tau= \left\{ \begin{array}{lr}
 \tau_{V} (\lambda /5500 \mathrm{\AA})^{-0.7} , & \mbox (\mathrm{t} \leq 10^{7} \mathrm{yrs})\\
\mu \tau_{V}(\lambda /5500 \mathrm{\AA})^{-0.7}, & \mbox (\mathrm{t} > 10^{7} \mathrm{yrs})
\end{array}\right. 
\end{equation}

for a grid of values: $\tau_{V}=0, 0.3, 0.6, 0.9, 1.2, 1.5$ and $\mu=0, 0.3, 0.5, 1.0$. The ISM contributes only a fraction, $\mu$, of the total attenuation. We use the Chabrier IMF, which produces results similar to the Kroupa IMF. From the UV luminosity, we calculate the star formation rate using Eq. 1. To simulate the radio emission we convert the supernova rate into a star formation rate using \citet{condon92}. However, we apply two additional conversion factors. First, the \citet{condon92} formalism uses a Miller-Scalo ({\em MS}) IMF. We convert to the Chabrier ({\em chab}) IMF in the following way:
 
\begin{eqnarray}
\mathrm{SFR(M>5M}_{\odot})_{chab}&=&\mathrm{SFR(M>5M}_{\odot})_{MS}\frac{\int_{5\mathrm{M}_{\odot}}^{100\mathrm{M}_{\odot}} \mathrm{M}^{-1.3}\,d\mathrm{M}}{\int_{\mathrm{5M}_{\odot}}^{\mathrm{100M}_{\odot}} \mathrm{M}^{-1.5}\,d\mathrm{M}} \\
&=& 1.75\mathrm{~SFR(M>M}_{\odot})_{MS}
 \end{eqnarray}
  Second, to find the SFR that includes all, not just the high mass (M $>$ 5M$_{\odot}$) stars, we correct by the factor
 \begin{eqnarray}
\mathrm{SFR}&=&SFR(M>5M_{\odot})_{chab}\frac{\int_{0.1\mathrm{M}_{\odot}}^{\mathrm{100M}_{\odot}}\mathrm{M}\Psi(\mathrm{M})\,d\mathrm{M}}{\int_{\mathrm{5M}_{\odot}}^{\mathrm{100M}_{\odot}} \mathrm{M}^{-1.5}\,d\mathrm{M}}\\
&=&2.44\mathrm{~SFR(M>5M}_{\odot})_{chab}
 \end{eqnarray}
where 
 \begin{equation}
  \Psi(\mathrm{log M}) \propto \left\{ \begin{array}{lr}
e^{-(\mathrm{log M} - \mathrm{log M}_{c})^2/2\sigma^2}, & \mbox (\mathrm{M} \le 1\mathrm{M}_{\odot}) \\
M^{-1.3}, & \mbox (\mathrm{M} > 1\mathrm{M}_{\odot})
\end{array}\right. 
 \end{equation}
 where M$_c=0.08$M$_{\odot}$ and $\sigma=0.69$. In Fig. \ref{sfrcheck}, we show how well the derived UV and radio SFR match the model SFR (these are unattenuated SFR). Since the conversions from luminosity or supernova rates to SFR were derived for constant star formation at solar metallicity, we see that in the Z=0.02, constant star formation scenario, radio and UV SFRs match the model SFR best. Furthermore, the conversion from UV luminosity to SFR overestimates the UV SFR for t$\gtrsim10^{7}$ years for the exponential decay scenario for all metallicity cases -- demonstrating the limitations of determining SFRs from UV luminosities if the SF history is unknown.  We also show how the ratio of radio SFR to unattenuated UV SFR changes with time for the exponentially decaying star formation model (at solar metallicity) in Fig. \ref{modage}. For t$<10^{7}$ yrs,  the radio SFR is less than the unattenuated UV SFR. 
 
Returning to the assertion that supercompact UVLGs may resemble the physical scenario where the majority of stars are within a ``birth cloud'', we compare the continuum and line attenuations in Fig. \ref{radxtau}. The top shows the histograms of the ratio of continuum to line attenuation for the comparison samples, which peak around 0.4 (similar to the results by \cite{calzetti94}). The supercompact UVLGs have a broader distribution, with a larger percentage of galaxies with higher continuum to line attenuation ratios.  We overplot two theoretical curves for the line versus continuum attenuation on our data in Fig. \ref{radxtau} (bottom). For our model data, the line attenuation is calculated from Eq. 6 where $\tau_V$ is varied for our model as described earlier. Similarly, for the continuum attenuation, the orange line marks the $t>10^{7}$yrs case for $\mu =0.3$, and the blue line gives the $\leq 10^{7}$, $\mu=1.0$ case, both at $\lambda=1500\mathrm{\AA}$. We acknowledge that galaxies have far more complicated attenuations than described by these two curves. However, the location of the supercompact UVLGs on this plot agrees with the interpretation that most of the stars in these galaxies are still enclosed in ``birth clouds''. Furthermore, Fig. \ref{radxtau} displays the lower limit in continuum attenuation, because the radio SFR underestimates the total SFR for t$<10^{7}$ yr (see Fig. \ref{modage}). 

\begin{figure}[tbp]
   \vspace{0.0in}
   \centering
   \includegraphics[height=6.5in]{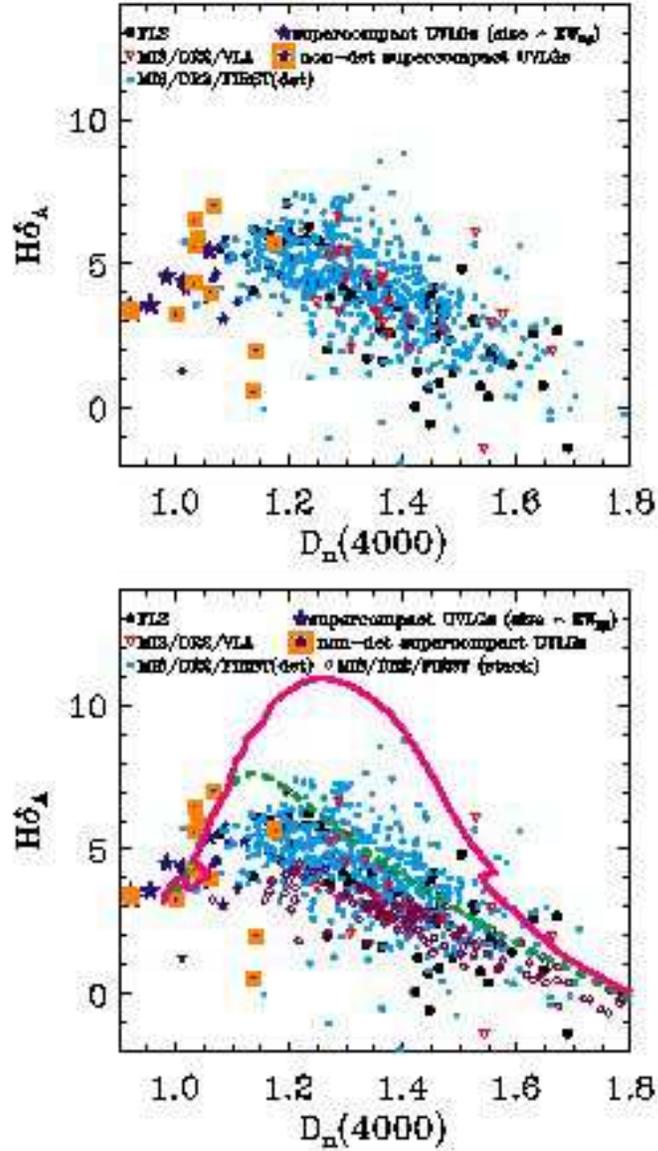} 
   \caption{The 4000$\mathrm{\AA}$~break index vs. the Balmer absorption line strength. The typical galaxies (black dots and inverted red triangles, as described in Fig 5) have decreasing H$\delta_{A}$, indicating the expiration of stars with strong Balmer absorption, with increasing D$_n$(4000). Supercompact UVLGs (shown as blue stars) may have undergone a very recent burst. The size of the stars reflect the H$\beta$ equivalent width, another indicator of recent to past star formation. In the bottom plot, we include the stacked MIS/DR2/FIRST galaxies and the models. The magenta solid line refers to an instantaneous burst of star formation with solar metallicity, whereas the green dotted line demonstrates exponentially decaying star formation with a decay constant of 2 Gyr and solar metallicity.\label{hdad4n}}
   \end{figure}

Approaching the timescale issue of the supercompact UVLGs from a different angle, we introduce a few other star formation history indicators. D$_n$(4000) measures the ratio of average flux densities (F$_{\nu}$) between two narrow wavelength regions: 4000-4100$\mathrm{\AA}$ and 3850-3950$\mathrm{\AA}$ \citep{balogh, bruzual}. The latter region contains an accumulation of absorption features from ionized metals, creating a sudden discontinuity around 4000$\mathrm{\AA}$. However, hotter stars can multiply ionize the metals, which reduces the opacity for this wavelength region and thereby eliminates the break. Young stellar populations lead to small D$_n$(4000), while old, metal-rich galaxies have high D$_n$(4000) values. The H$\delta$ absorption strength (H$\delta_{A}$) also estimates the stellar population age. A-type stars have the most prominent Balmer absorption features. O and B stars dilute the absorption lines, since they dominate the spectrum yet have weak absorption. After the O and B stars perish ($\sim 0.1 - 1$ Gyr) H$\delta_{A}$ peaks; as the A stars finish their evolution the H$\delta$ absorption strength decreases. The H$\beta$ equivalent width (EW(H$\beta$)) takes into account both the instantaneous star formation rate and the entire star formation history, and effectively measures the strength of the recent star formation activity compared to past. In Fig. \ref{hdad4n}, we combine these different measures, plotting the D$_n$(4000) versus H$\delta_{A}$, and for the supercompact UVLGs the symbol size indicates EW(H$\beta$). The dashed green line and solid magenta line follow an exponentially decaying (with a time constant of 2 Gyr) and instantaneous burst star formation history, respectively (they both increase in time from low D$_n$(4000) to higher D$_n$(4000)). Given their low D$_n$(4000) and low H$\delta_{A}$, the spectra of supercompact UVLGs are dominated by recent star formation. 

\begin{figure}[tbp]
   \vspace{0.0in}
   \centering
   \includegraphics[height=5in]{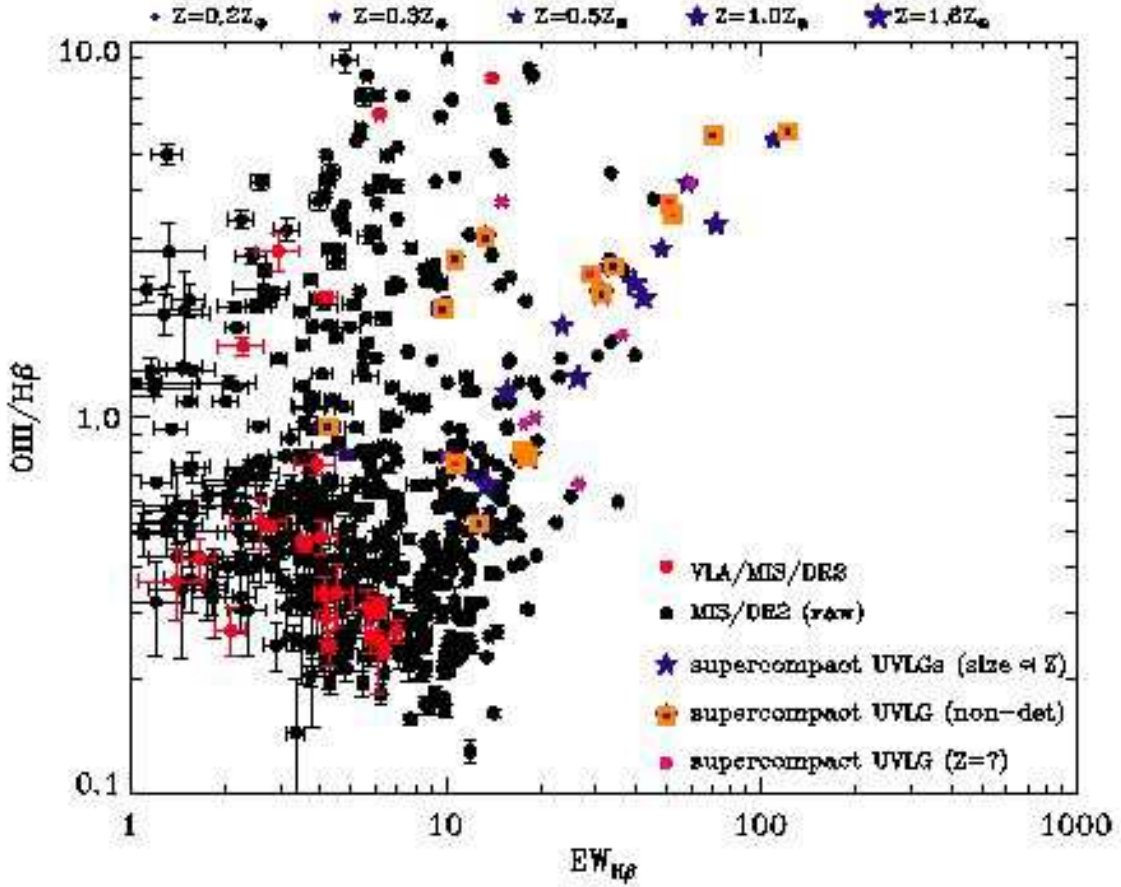} 
   \caption{The OIII/H$\beta$ versus EW(H$\beta$) can segregate star formation histories. Bursts, on underlying older stellar populations have high OIII/H$\beta$, but low EW(H$\beta$) while single, recent bursts have high OIII/H$\beta$ and  EW(H$\beta$). The supercompact UVLGs (blue) seem to be the latter, although low metallicity may contribute to high OIII/H$\beta$ values more than burst history. To qualitatively test this point, we indicate metallicity by symbol size for the supercompact UVLGs (the magenta points have undetermined metallicities) and a rough guide relating the symbol size to metallicity is shown at the top of panel($Z_\odot=8.69$). The black dots are {\em unstacked} MIS/DR2/FIRST galaxies. \label{rg1}}
      \end{figure}

\citet{RG06} study the radio properties of HII galaxies along with the H$\beta$ equivalent width (EW(H$\beta$)) and [OIII]$\lambda 5007\mathrm{\AA}$ to H$\beta$ ratio. [OIII] is excited by massive stars, and strong [OIII] emission lines indicate early phases of star formation activity. While the EW(H$\beta$) gauges present to past star formation, the [OIII]/H$\beta$ ratio detects recent starbursts. Therefore, single bursts would have high EW(H$\beta$) and high [OIII]/H$\beta$, while bursts on underlying older stellar populations might have low EW(H$\beta$) but high [OIII]/H$\beta$. Fig.\ref{rg1} shows that supercompact UVLGs fall in the region with highest EW(H$\beta$) and highest [OIII]/H$\beta$-- therefore, not only have they had recent bursts (as established already by their UV SFR), but their present to past star formation rates exceed that observed in other star-forming galaxies. However, we note one important caveat: metallicity affects the [OIII]/H$\beta$ measure. The size of the symbols in Fig.\ref{rg1} indicates metallicity for the supercompact UVLGs. Lower metallicity results in higher gas temperatures, making it easier to collisionally excite the [OIII] line. Given that supercompact UVLGs are metal-poor \citep{Choopes}, we are less certain that the [OIII]/H$\beta$ indicates starbursts. Since the cause of the high [OIII]/H$\beta$ values is uncertain for the these galaxies, we cannot confidently date the star formation, but compared to other galaxies supercompact UVLGs show evidence of recent activity. 

HII galaxies have similar [OIII]/H$\beta$ properties \citep{RG06}, however these galaxies have FUV surface brightnesses (or star formation rates per unit area) an order of magnitude lower than supercompact UVLGs and FUV luminosities 2 orders of magnitude lower than the supercompact UVLGs \citep{Choopes}. We briefly consider the connection between supercompact UVLGs and HII galaxies. The spectra of HII galaxies resemble HII regions, where embedded young stars dominate the emission. The radio emission in these galaxies is dominated by the thermal, rather than synchrotron, contribution. Since the thermal bremsstrahlung spectrum is characterized by a $\nu^{-0.1}$ power law, while synchrotron emission follows $\nu^{-0.8}$, the spectral index ($\alpha$, where S$_\nu \propto \nu^{\alpha}$) can discern which mechanism is responsible for the radio emission. Furthermore, an accurate measure of the SFR from the radio data depends on understanding the contribution of thermal versus non-thermal; current conversions assume that 90\% of the radio is non-thermal at 20 cm. We have acquired 6 cm radio data, in order to determine spectral indices for the supercompact UVLGs, and our preliminary results indicate that the non-thermal contribution dominates the spectrum, consistent with the normal FIR/radio flux ratios that we measure. 

We have outlined two explanations for why the radio luminosities (for given UV luminosities) might be low for the supercompact UVLGs: less dust attenuation and recent starburst activity. If we assume that the radio-IR correlation implies consistent SFRs, then the issue of recent star formation is less relevant. Although we will investigate this further in a future paper, our IR data suggests low dust attenuation is the more plausible explanation for our observations. 

\section{Conclusions}
\begin{figure}[tbp]
   \vspace{0.0in}
   \centering
   \includegraphics[height=7in]{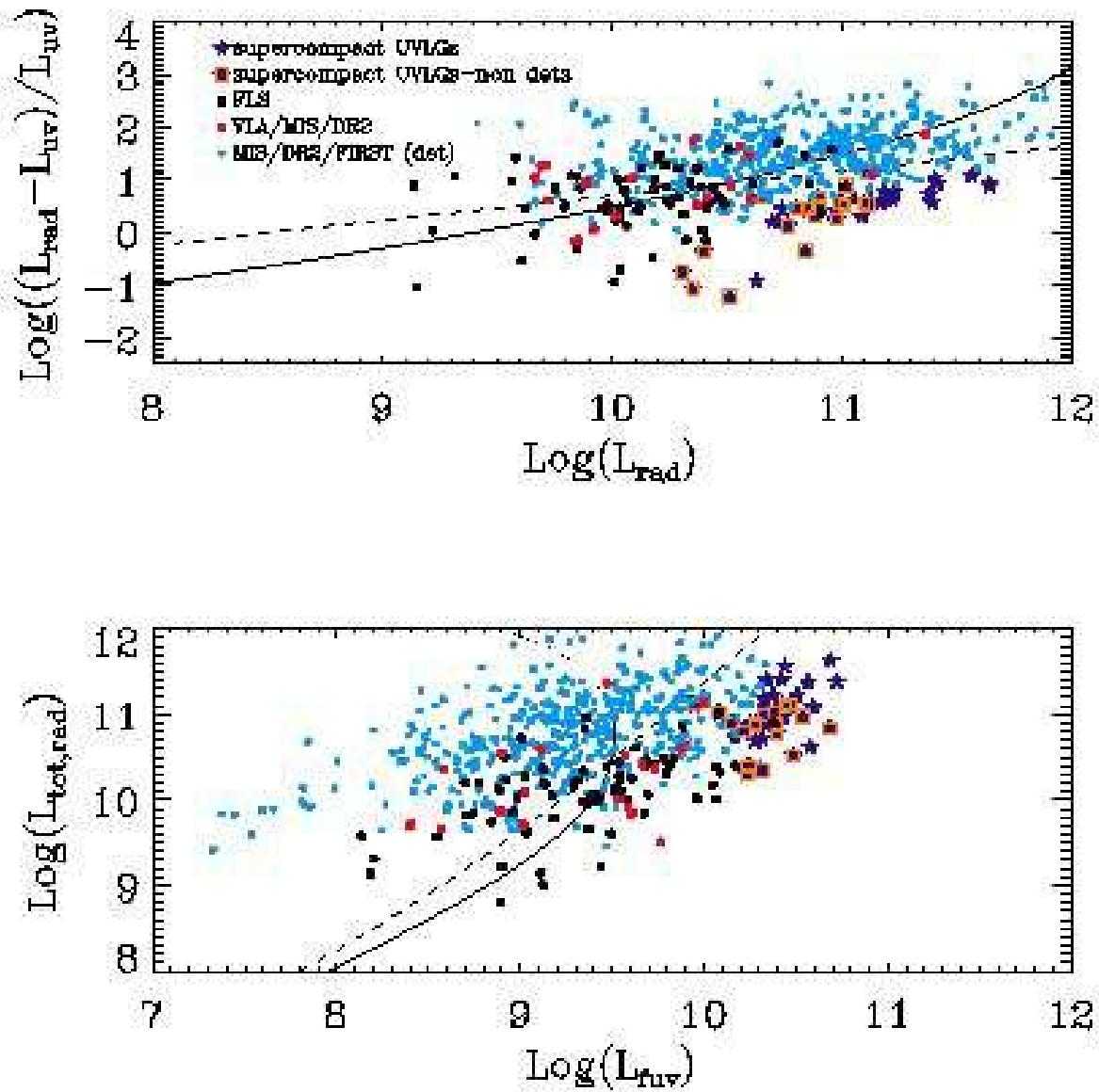} 
   \caption{Radio luminosity versus IR/UV luminosity (top) with fits from \citet{Martin05}(solid) and \citet{Wang96} (dashed). UV luminosity vs. radio luminosity (bottom). The lines show the radio luminosity derived from \citet{Martin05}(solid) and \citet{Wang96} (dashed). The FLS and VLA/MIS/DR2 galaxies have scatter both above and below the fits, while the supercompact UVLGs all have radio luminosities that fall below the relation. \label{chris}}
\end{figure}

\begin{figure}[tbp]
   \vspace{0.0in}
   \centering
   \includegraphics[height=3.5in]{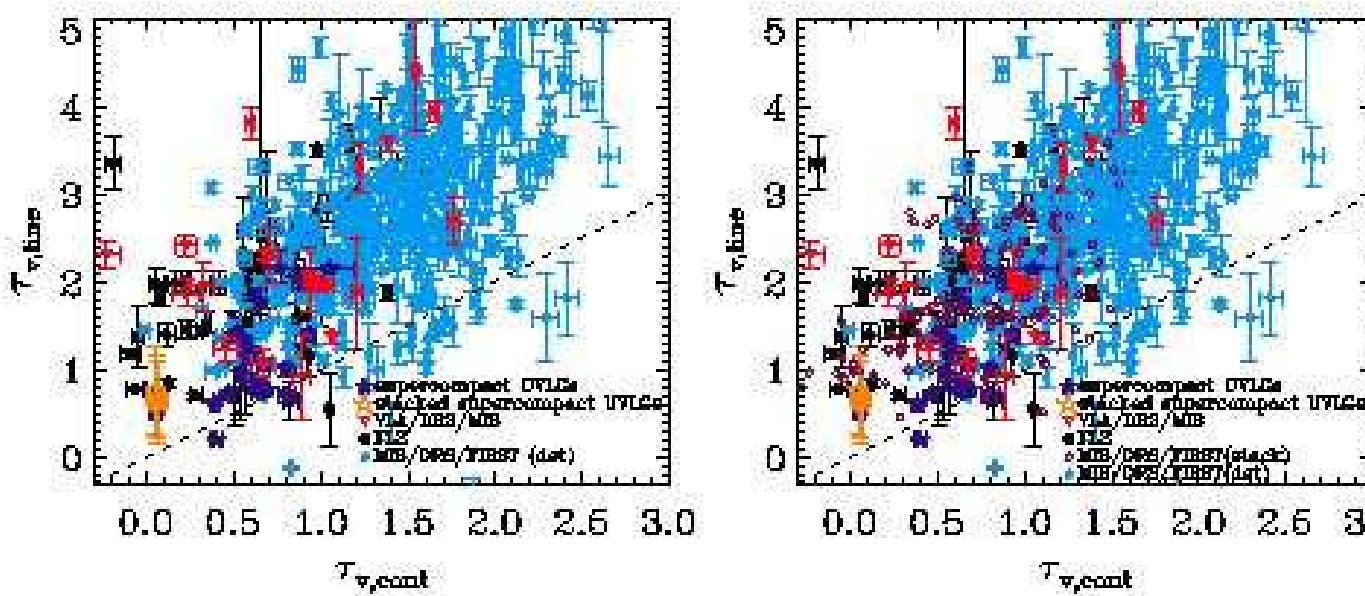}
   \caption{Attenuation in the V magnitude of the $H\alpha/H\beta$ lines ($\tau_{V,line}$) versus the continuum ($\tau_{V,cont}$), measured from radio-to-UV SFR ratio (See Eqns 9 and 10): The samples are the same as in Fig 4. The left panels exclude the stacked MIS/DR2/FIRST galaxies to visually simplify the plot, while the panels on the right display the full set. The dashed line shows where $\tau_{V,line}=\tau_{V, cont}$. \label{tau}}
   \end{figure}

\begin{figure}[tbp]
   \vspace{0.0in}
   \centering
   \includegraphics[height=7in]{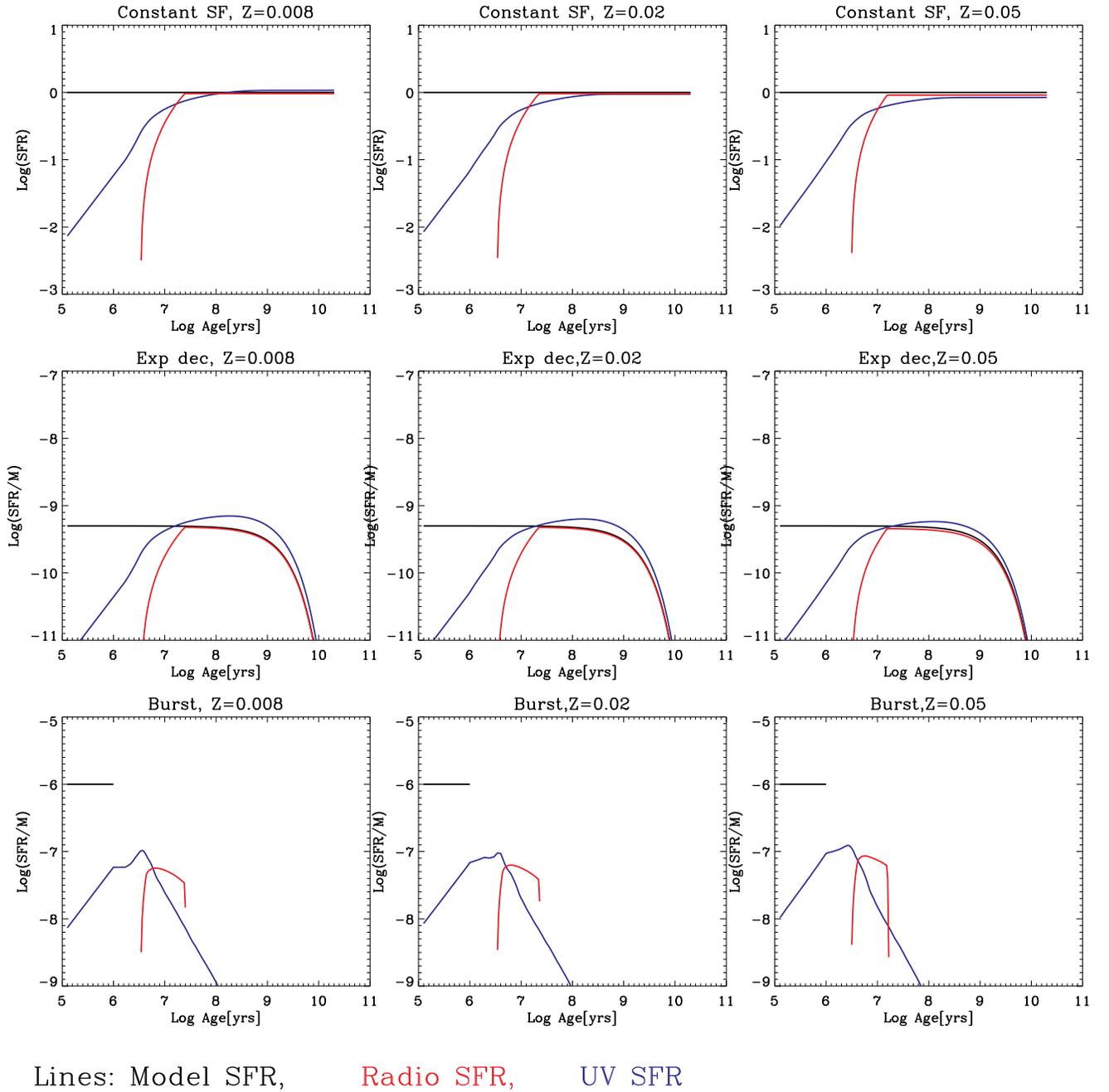} 
   \caption{Comparing different star formation indicators in models: Radio SFR (red), UV SFR (blue) and model input(black). The rows display different star formation scenarios: constant, exponentially declining, and burst (top, middle, bottom), while the columns illustrate the effect of different metallicities: Z=0.008, 0.02, and 0.05 (left, middle, and right). \label{sfrcheck}}
   \end{figure}

\begin{figure}[tbp]
   \vspace{0.0in}
   \centering
   \includegraphics[height=7in]{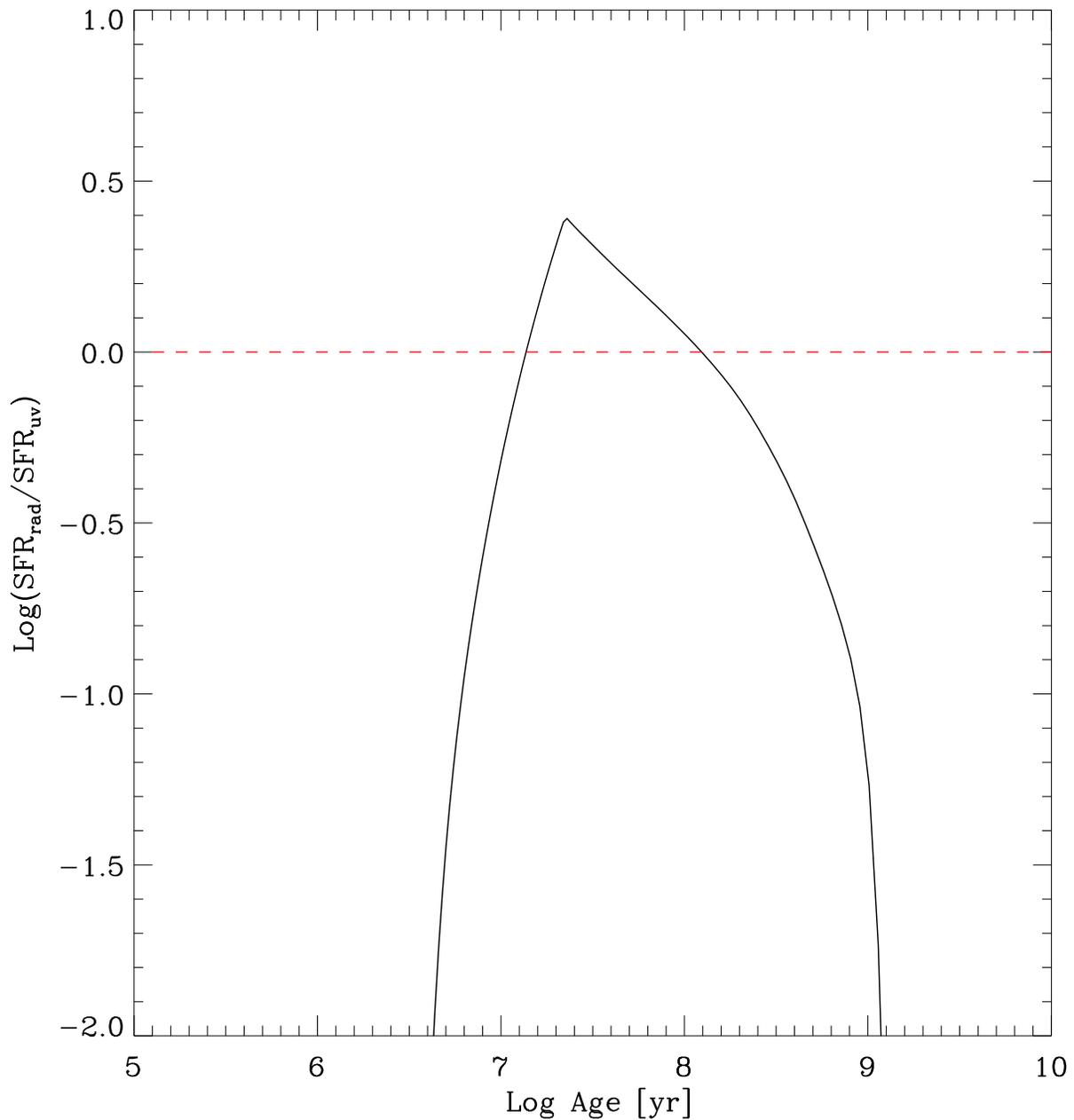} 
   \caption{The radio to UV (unattenuated) SFR ratio versus Log(Age) for the exponentially decaying star formation scenario with solar metallicity and decay constant of 100 Myr.  The red dashed line marks where the radio SFR = UV SFR. The plot demonstrates that for $t<10^{7}$ yr, the unattenuated UV SFR is greater than the radio SFR.\label{modage}}
   \end{figure}

\begin{figure}[tbp]
   \vspace{0.0in}
   \centering
    \includegraphics[height=2.5in]{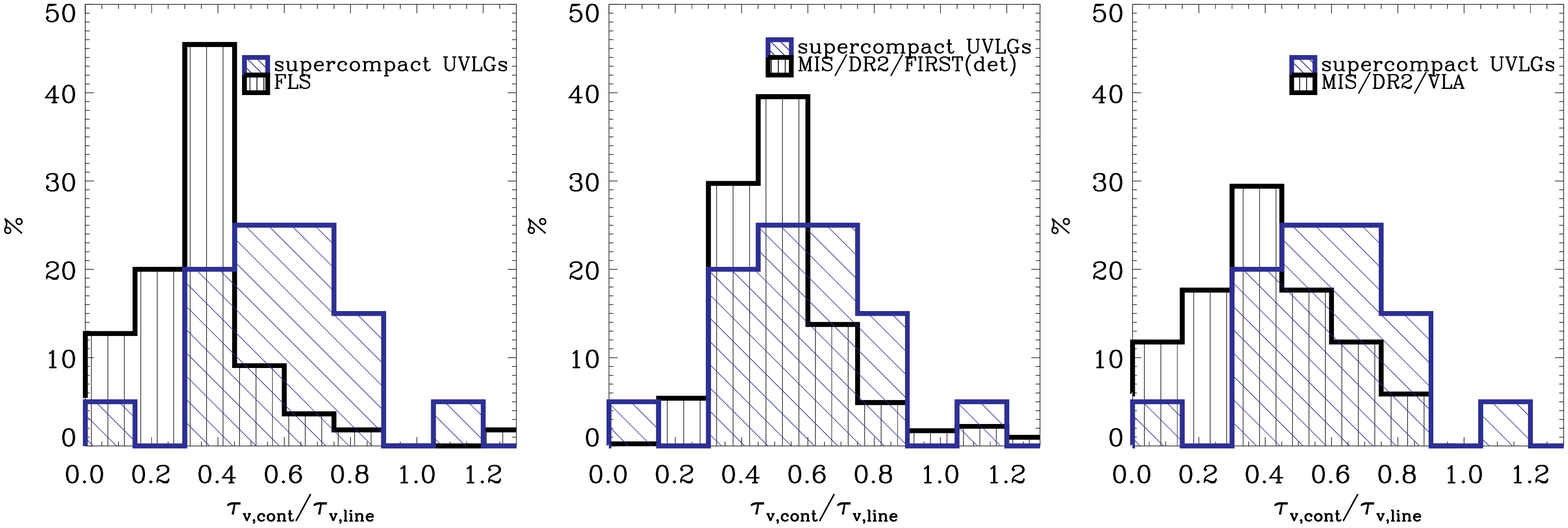} 
   \includegraphics[height=3in]{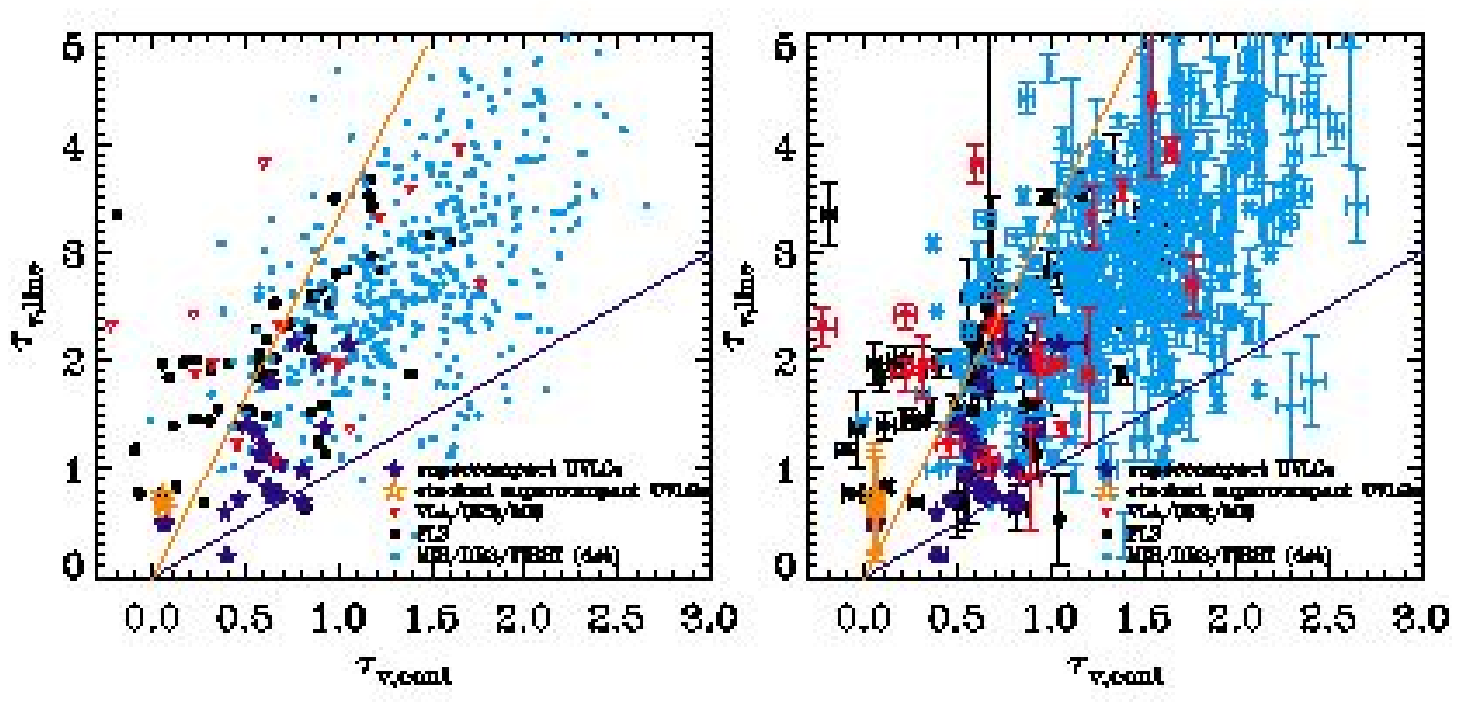} 
   \caption{Comparing two different measures of dust. $\tau_{V,cont}\equiv$Ln($SFR_{radio}/ SFR_{uv}$)/2.483 measures attenuation of the continuum, while $\tau_{V,line}\equiv ln[\frac{H\alpha}{H\beta}\frac{1}{2.88}]$*4.84 measures attenuation in spectral lines. The histograms (top) show that the ratio of continuum to line attenuation peaks for normal galaxies (FLS, MIS/DR2/FIRST(dets), and MIS/DR2/VLA from left to right, respectively) $\sim 0.4$, while the supercompact UVLGs have a broader distribution with a larger percentage of galaxies with higher ratios. In the lower panel, the orange line shows the theoretical curve for a galaxy where the attenuation in the continuum is 30\% of the attenuation in the line ($\mu=$0.3), while the blue line shows the case where the line and continuum are attenuated equally (all the stars are still in birth clouds, $\mu=$1)). For a clearer view, the left panel excludes error bars and the MIS/DR2/FIRST stacked galaxies, while the right panel shows the full set with errors. \label{radxtau}}
   \end{figure}

Our study has combined the 1.4 GHz radio data with UV, several spectral measures, and IR (in some cases) to understand the dust attenuation properties and star formation histories of supercompact UVLGs-- compared to typical, local galaxies. For galaxies with no dust attenuation, the UV SFR should measure the total SFR; for galaxies with stellar populations older than $\sim10^{7}-10^{7.5}$ yrs the 1.4 GHz SFR should measure the total SFR (see Figs. \ref{sfrcheck} and \ref{modage}). In light of these assumptions, we note the following results:
 
\begin{enumerate}
\item{The supercompact UVLGs are consistent with radio-IR relationship found by \citet{Yun}. However, to compare the IR SFR with radio SFR, we need to complete further analysis of the IR data. This result suggests that the radio continuum can be used as a proxy for the FIR and can (in conjunction with the FUV) be used to determine the SFR and FUV extinction.}
\item {We derive that the average $L_{rad}$=5.8$\times 10^{22}$ W Hz$^{-1}$, which is comparable to the value found by \citet{Reddy}: 5.9$\times 10^{22}$ W Hz$^{-1}$ for stacked 1.5 $\leq z \leq$ 3.0 LBGs. The implied SFRs are in the range of 10-100M$_\odot$ yr$^{-1}$, similar to LBGs.}
\item{For a given FUV luminosity, the radio luminosity appears to be lower than what is expected using relations derived by \cite{Wang96} and \cite{Martin05}. While other samples of galaxies have points scattered around the relation, the supercompact UVLGs have $L_{rad}$ that fall below the relation (see Fig. \ref{chris}). }
\item{Our average SFR$_{rad}/$SFR$_{UV} \sim$ 4.7, consistent with LBGs \citep{Reddy}. Compared to typical local galaxies with such high SFRs (i.e.LIRGs), the supercompact UVLGs have significantly less FUV attenuation. The combination of modest FUV attenuation with high SFR is similar to what is observed in LBGs.}
\item{Supercompact UVLGs have lower line attenuation, but comparable continuum attenuation to other typical galaxies (Fig. \ref{tau}). We compute the line attenuations from Balmer decrements ($\tau=$ln$(\frac{H\alpha}{H\beta}\frac{1}{2.88})$). The SFR$_{rad}/$SFR$_{UV}$ quantity estimates the continuum attenuation.  We consider the effect of birth clouds on line versus continuum attenuation to derive simple theoretical curves, which we overplot with the data in Fig. \ref{radxtau}. The location of the supercompact UVLGs on the $\tau_{V,line}-\tau_{V,cont}$ plot is consistent with the interpretation that stars in supercompact UVLGs are embedded in ``birth clouds''.}
\item{Compared to our set of comparison galaxies, defined in Sections 2.2 through 2.4, supercompact UVLGs have: lower D$_n$(4000) and lower H$\delta_{A}$ (implying the presence of young O and B stars), higher EW($H\beta$) and higher [OIII]/H$\beta$ (suggesting that most of the stars were formed recently). }
\end{enumerate}

These results consistently indicate that supercompact UVLGs are low dust systems with recent star formation, and reinforce their connection with high-redshift LBGs. 

\url[]{GALEX} (Galaxy Evolution Explorer) is a NASA Small Explorer, launched in April 2003. We gratefully acknowledge NASAÕs support for construction, operation, and science analysis for the GALEX mission, developed in cooperation with the Centre National d'Etudes Spatiales of France and the Korean Ministry of Science and Technology. We thank Michael Blanton for access to the IDL kcorrect (version 4.1.4) analysis package. We recognize that funding for the SDSS archive has been provided by the Alfred P. Sloan Foundation, the Participating Institutions, the National Aeronautics and Space Administration, the National Science Foundation, the U.S. Department of Energy, the Japanese Monbukagakusho, and the Max Planck Society.


\begin{deluxetable}{lccccccc}

\setlength{\tabcolsep}{0.05in}
\tabletypesize{\footnotesize}
\tablecolumns{8} 
\tablewidth{0pc} 
\tablecaption{Supercompact UVLGs} 
\tablehead{ 
\colhead{SDSS ID}  & \colhead{$z$} & \colhead{S$_{70\mu m}$} & \colhead{S$_{1.4\mathrm{GHz}}$\tablenotemark{a}} &\colhead{L$_{1.4\mathrm{GHz}}$\tablenotemark{b}} & \colhead{SFR$_{1.4\mathrm{GHz}}$} &  \colhead{L$_{\mathrm{FUV}}$\tablenotemark{b}} & \colhead{SFR$_{\mathrm{FUV}}$}\\
\colhead{} & \colhead{} & \colhead{(mJy)} &\colhead{($\mu$Jy)} & \colhead{( $10^{22}$ W Hz$^{-1}$)} & \colhead{(M$_\odot$ yr$^{-1}$)} & \colhead{( $10^{22}$ W Hz$^{-1}$)} & \colhead{(M$_\odot$ yr$^{-1}$)}
}
\startdata 

 SDSS~J205000.00$+$003124.7  &   0.164  &  \nodata  &  $<$   324  &  $<$   2.3  &  $<$   8.5   &    0.15$\pm$  0.07  &     1.4$\pm$   0.6\\
 SDSS~J221155.99$-$093223.1  &   0.209  &  \nodata  &     708$\pm$    72  &     8.6$\pm$   0.9  &    31.5$\pm$   3.2  &    0.47$\pm$  0.07  &     4.3$\pm$   0.6\\
 SDSS~J232539.22$+$004507.2  &   0.277  &  $<$    17   &     338$\pm$    54  &     7.7$\pm$   1.2  &    28.2$\pm$   4.5  &    0.41$\pm$  0.13  &     3.8$\pm$   1.2\\
 SDSS~J232624.84$+$134206.4  &   0.207  &  \nodata  &  $<$   295  &  $<$   3.5  &  $<$  12.9   &    0.95$\pm$  0.14  &     8.8$\pm$   1.3\\
 SDSS~J001009.97$-$004603.6  &   0.243  &  $<$    12  &  $<$   215  &  $<$   3.7  &  $<$  13.4   &    0.37$\pm$  0.13  &     3.5$\pm$   1.2\\
 SDSS~J015028.40$+$130858.3  &   0.147  &     477$\pm$     7  &    1500$\pm$    73  &     8.4$\pm$   0.4  &    30.9$\pm$   1.5  &    0.63$\pm$  0.08  &     5.9$\pm$   0.7\\
 SDSS~J015125.97$+$132510.8  &   0.243  &  \nodata  &  $<$   294  &  $<$   5.0  &  $<$  18.2   &    0.42$\pm$  0.13  &     3.9$\pm$   1.2\\
 SDSS~J021348.53$+$125951.4  &   0.219  &  \nodata  &     896$\pm$    68  &    12.1$\pm$   0.9  &    44.3$\pm$   3.4  &    0.72$\pm$  0.22  &     6.7$\pm$   2.1\\
 SDSS~J032845.99$+$011150.8  &   0.142  &  $<$    11   &  $<$   240  &  $<$   1.3  &  $<$   4.6  &    0.34$\pm$  0.12  &     3.2$\pm$   1.1\\
 SDSS~J035733.99$-$053719.7  &   0.204  &      49$\pm$     6  &     444$\pm$    85  &     5.1$\pm$   1.0  &    18.7$\pm$   3.6  &    0.43$\pm$  0.16  &     4.0$\pm$   1.5\\
 SDSS~J081523.39$+$500414.6  &   0.164  &  \nodata  &     672$\pm$    63  &     4.9$\pm$   0.5  &    17.7$\pm$   1.7  &    0.54$\pm$  0.07  &     5.0$\pm$   0.7\\
 SDSS~J101211.18$+$632503.6  &   0.246  &  $<$    16   &  $<$   230  &  $<$   4.0  &  $<$  14.7  &    0.28$\pm$  0.10  &     2.6$\pm$   0.9\\
 SDSS~J101741.02$+$510438.4  &   0.213  &  \nodata  &  $<$   234  &  $<$   3.0  &  $<$  10.8   &    0.49$\pm$  0.07  &     4.5$\pm$   0.7\\
 SDSS~J102959.95$+$482937.9  &   0.232  &  \nodata  &     853$\pm$    56  &    13.2$\pm$   0.9  &    48.1$\pm$   3.2  &    0.52$\pm$  0.10  &     4.8$\pm$   0.9\\
 SDSS~J105145.51$+$660621.3  &   0.170  &  \nodata  &     363$\pm$    65  &     2.8$\pm$   0.5  &    10.2$\pm$   1.8  &    0.29$\pm$  0.04  &     2.7$\pm$   0.4\\
 SDSS~J113303.79$+$651341.3  &   0.241  &  $<$    16   &  $<$   285  &  $<$   4.8  &  $<$  17.5  &    0.67$\pm$  0.11  &     6.2$\pm$   1.0\\
 SDSS~J113947.89$+$630911.3  &   0.246  &  \nodata  &  $<$   300  &  $<$   5.2  &  $<$  19.1   &    0.47$\pm$  0.11  &     4.4$\pm$   1.1\\
 SDSS~J135355.90$+$664800.5  &   0.198  &      70$\pm$     4  &     567$\pm$    67  &     6.2$\pm$   0.7  &    22.5$\pm$   2.7  &    0.77$\pm$  0.10  &     7.1$\pm$   0.9\\
 SDSS~J223429.58$-$092452.9  &   0.246  &  \nodata  &  $<$   373  &  $<$   6.5  &  $<$  23.9   &    0.17$\pm$  0.08  &     1.6$\pm$   0.8\\
 SDSS~J230703.76$+$011311.1  &   0.126  &  \nodata  &     664$\pm$    76  &     2.7$\pm$   0.3  &     9.8$\pm$   1.1  &    0.41$\pm$  0.05  &     3.8$\pm$   0.4\\
 SDSS~J214500.25$+$011157.5  &   0.204  &      23$\pm$     5  &     341$\pm$    69  &     4.0$\pm$   0.8  &    14.5$\pm$   2.9  &    0.37$\pm$  0.08  &     3.4$\pm$   0.7\\
 SDSS~J231812.99$-$004126.1  &   0.252  &      65$\pm$     4  &     679$\pm$    72  &    12.5$\pm$   1.3  &    45.7$\pm$   4.8  &    1.03$\pm$  0.12  &     9.6$\pm$   1.1\\
 SDSS~J001054.85$+$001451.3  &   0.243  &  111$\pm$    5  &     753$\pm$    74  &    12.9$\pm$   1.3  &    47.1$\pm$   4.6  &    0.43$\pm$  0.09  &     4.0$\pm$   0.8\\
 SDSS~J004447.33$+$152911.7  &   0.227  &  \nodata  &     462$\pm$    78  &     6.8$\pm$   1.1  &    24.8$\pm$   4.2  &    0.53$\pm$  0.09  &     4.9$\pm$   0.8\\
 SDSS~J005527.46$-$002148.7  &   0.167  &     163$\pm$     6  &     542$\pm$    72  &     4.1$\pm$   0.5  &    14.9$\pm$   2.0  &    0.45$\pm$  0.08  &     4.2$\pm$   0.8\\
 SDSS~J024529.55$-$081637.7  &   0.196  &  \nodata  &     373$\pm$    67  &     3.9$\pm$   0.7  &    14.4$\pm$   2.6  &    0.35$\pm$  0.06  &     3.3$\pm$   0.5\\
 SDSS~J040208.86$-$050642.0  &   0.139  &  $<$    20   &  $<$   225  &  $<$   1.1  &  $<$   4.1  &    0.40$\pm$  0.07  &     3.7$\pm$   0.6\\
 SDSS~J092336.45$+$544839.2  &   0.222  &      45$\pm$     4  &   $<$   276  &  $<$   3.9  &  $<$  14.1  &    0.49$\pm$  0.08  &     4.6$\pm$   0.7\\
 SDSS~J210358.74$-$072802.4  &   0.137  &     421$\pm$     6  &    3840$\pm$    48  &    18.6$\pm$   0.2  &    68.1$\pm$   0.9  &    0.54$\pm$  0.24  &     5.0$\pm$   2.2\\
 SDSS~J004054.32$+$153409.6  &   0.283  &  25$\pm$   5 &  $<$   220  &  $<$   5.3  &  $<$  19.3   &    0.24$\pm$  0.09  &     2.2$\pm$   0.9\\
 SDSS~J020356.91$-$080758.5  &   0.189  &  \nodata  &  $<$   675  &  $<$   6.6  &  $<$  24.0   &    0.57$\pm$  0.07  &     5.3$\pm$   0.7\\
 SDSS~J005439.79$+$155446.9  &   0.237  &  \nodata  &  $<$   200  &  $<$   3.2  &  $<$  11.7   &    0.32$\pm$  0.09  &     3.0$\pm$   0.8\\
 SDSS~J080232.34$+$391552.6  &   0.267  &      85$\pm$     6  &     352$\pm$    52  &     7.4$\pm$   1.1  &    27.1$\pm$   4.0  &    0.60$\pm$  0.11  &     5.5$\pm$   1.1\\
 SDSS~J082001.72$+$505039.1  &   0.217  &      68$\pm$     5  &     433$\pm$    61  &     5.8$\pm$   0.8  &    21.1$\pm$   3.0  &    0.42$\pm$  0.09  &     3.9$\pm$   0.8\\
 SDSS~J083803.72$+$445900.2  &   0.143  &      18$\pm$     5  &   $<$   750  &  $<$   4.0  &  $<$  14.7  &    0.37$\pm$  0.04  &     3.5$\pm$   0.4\\
 SDSS~J093813.49$+$542825.0  &   0.102  &     123$\pm$     5  &   $<$   635  &  $<$   1.7  &  $<$   6.0  &    0.60$\pm$  0.04  &     5.6$\pm$   0.3\\
 SDSS~J102613.97$+$484458.9  &   0.160  &      42$\pm$     4  &     369$\pm$    51  &     2.5$\pm$   0.3  &     9.2$\pm$   1.3  &    0.37$\pm$  0.02  &     3.4$\pm$   0.1\\
 SDSS~J143417.15$+$020742.5  &   0.180  &      76$\pm$     6  &     459$\pm$    56  &     4.0$\pm$   0.5  &    14.8$\pm$   1.8  &    0.40$\pm$  0.06  &     3.7$\pm$   0.5\\
 SDSS~J080844.26$+$394852.3  &   0.091  &     156$\pm$     7  &   $<$   500  &  $<$   1.0  &  $<$   3.7  &    0.34$\pm$  0.04  &     3.1$\pm$   0.4\\
 SDSS~J092159.38$+$450912.3  &   0.235  &     228$\pm$     8  &    1410$\pm$    51  &    22.3$\pm$   0.8  &    81.5$\pm$   2.9  &    0.96$\pm$  0.17  &     8.9$\pm$   1.6\\
 SDSS~J092600.40$+$442736.1  &   0.181  &      21$\pm$     4  &     242$\pm$    42  &     2.1$\pm$   0.4  &     7.8$\pm$   1.4  &    0.74$\pm$  0.07  &     6.8$\pm$   0.6\\
 SDSS~J124819.74$+$662142.6  &   0.260  &  \nodata  &  $<$   320  &  $<$   6.3  &  $<$  23.2   &    0.54$\pm$  0.12  &     5.0$\pm$   1.1\\
stack~1\tablenotemark{c} & 0.22 & \nodata &141$\pm$ 21 & 1.6$\pm$0.2 &5.8$\pm$ 1.1&0.45$\pm$0.09&4.2$\pm$0.8\\
stack~2\tablenotemark{c} & 0.22 & \nodata & 157$\pm$ 18 & 1.3$\pm$0.1  &4.7$\pm$ 0.9&0.45$\pm$0.10&4.2$\pm$1.0\\
stack\_all\tablenotemark{c}& 0.22 & \nodata & 149$\pm$ 15 & 1.4$\pm$0.1  & 5.2$\pm$0.7&0.45 $\pm$0.10&4.2$\pm$1.0\\
\enddata 

\tablenotetext{a}{$S_{1.4GHz}$ gives the measured (not k-corrected) value.}
\tablenotetext{b} {Luminosities were calculated using the following cosmology: H$_{0}$= 70 km s$^{-1}$Mpc$^{-1}$,
 $\Omega_{M}$=0.3, $\Omega_{\nu}$= 0.7. We compute exposure weighted averages from the FUV magnitudes.}
 \tablenotetext{c}{Stack 1 includes all non-detected galaxies from our first VLA observation. Stack 2 includes the non-detected galaxies from the second VLA observation. Finally, stack~all includes the galaxies from both stack1 and stack2. Details about the stacking process are discussed in the text. The redshift, radio flux density, radio and FUV luminosities and SFRs are all averaged values.}
\end{deluxetable} 

\begin{deluxetable}{clcccc}
\tablecolumns{6} 
\tablewidth{0pc} 
\tablecaption{Summary of Data Sets} 
\tablehead{ 
\colhead{Data Set} & \colhead{\#} & \colhead{Median z} & \colhead{FUV\tablenotemark{a}} & \colhead{Optical\tablenotemark{a}} & \colhead{Radio\tablenotemark{a}}
}
\startdata 
& & &\GALEX~MIS/AIS\tablenotemark{c} & SDSS DR2 & VLA\\
Supercompact UVLGs& 42 &0.21 & 1500 s(MIS)/100 s(AIS) & 54 s & 1 hr \\
 & & &$m_{FUV} < 23(MIS)/20.5(AIS)$ & $m_{r}<21$ & $S>350\mu$Jy \\
\tableline
 &  & &\GALEX & SDSS & VLA\\
FLS& 189&0.11& 2500 s & 54 s & 5 hr \\
 & & &$m_{FUV} < 24.7$ & $m_{r}<22.2$\tablenotemark{b} & $S>115\mu$Jy \\
\tableline
& & &\GALEX~MIS & SDSS DR2 & FIRST\tablenotemark{d}\\
MIS/DR2/FIRST & 21000 &0.08 & 1500 s & 54 s  & 3 min \\
 & & &$m_{FUV} < 23$ & $m_{r}<22.2$ & $S>500\mu$Jy \\
\tableline
 & & &\GALEX~MIS& SDSS DR2 & VLA\\
MIS/DR2/VLA & 24 & 0.07 & 1500 s & 54 s & 1 hr \\
 & & &$m_{FUV} < 23$ & $m_{r}<22.2$ & $S>350\mu$Jy \\

 \enddata 
\tablenotetext{a} {Source for data, exposure time, sensitivity (5$\sigma$ point source).}
\tablenotetext{b} {The sensitivity limits for the entire filter set: $m_{u}<22.0, m_{g}<22.2, m_{r}<22.2, m_{i} <21.3, m_{z}<20.5$}
\tablenotetext{c} {In cases where the galaxy is observed multiple times, we compute an exposure-weighted average FUV magnitude.}
\tablenotetext{d} {Table values reflect the raw (unstacked) data. Since the 119 separate redshift, $\tau$, and $L_{FUV}$ bins contain different numbers ($31<N<829$) of galaxies, the effective exposure times and sensitivities vary: 1.5 hr$<t_{exp}<$ 49 hr and 27 $\mu$Jy$<S<$740 $\mu$Jy. Errors determined by bootstrapping for these stacked images are 6.5 $\mu$Jy$<RMS<$42 $\mu$Jy.}

\end{deluxetable} 

\clearpage

   \end{document}